\documentclass{emulateapj}
\usepackage{natbib,graphicx,epstopdf}
\usepackage{subfigure}

\newcommand{\wspitzer}{\emph{Warm-Spitzer}}
\newcommand{\spitzer}{\emph{Spitzer}}
\newcommand{\kepler}{\emph{Kepler}}
\newcommand{\corot}{\emph{CoRoT}}

\newcommand{\kms}{\ensuremath{\rm km\,s^{-1}}}
\newcommand{\ms}{\ensuremath{\rm m\,s^{-1}}}
\newcommand{\gcmc}{\ensuremath{\rm g\,cm^{-3}}}
\newcommand{\rhk}{\ensuremath{R^{\prime}_{\rm HK}}}	
\newcommand{\logrhk}{\ensuremath{\log\rhk}}		
\newcommand{\sval}{\ensuremath{S_{\mbox{\scriptsize HK}}}}
\newcommand{\caii}{\ion{Ca}{2} H \& K}
\newcommand{\chisq}{\mbox{$\chi^2$}\,}

\newcommand{\teff}{\ensuremath{T_{\rm eff}}}
\newcommand{\logg}{\ensuremath{\log{g}}}
\newcommand{\vsini}{\ensuremath{v \sin{i}}}
\newcommand{\feh}{[Fe/H]}
\newcommand{\mh}{[M/H]}

\newcommand{\vrot}{\ensuremath{V_{\rm rot}}}

\newcommand{\rsun}{\ensuremath{R_\sun}}
\newcommand{\msun}{\ensuremath{M_\sun}}

\newcommand{\rstar}{\ensuremath{R_\star}}
\newcommand{\mstar}{\ensuremath{M_\star}}
\newcommand{\loggstar}{\ensuremath{\logg_\star}}

\newcommand{\rhostar}{\ensuremath{\rho_\star}}

\newcommand{\rpl}{\ensuremath{R_{\rm P}}}
\newcommand{\mpl}{\ensuremath{M_{\rm P}}}

\newcommand{\rhopl}{\ensuremath{\rho_{\rm P}}}
\newcommand{\loggpl}{\ensuremath{\logg_{\rm P}}}
\newcommand{\teq}{\ensuremath{T_{\rm eq}}}

\newcommand{\rjup}{\ensuremath{R_{\rm J}}}
\newcommand{\mjup}{\ensuremath{M_{\rm J}}}

\newcommand{\koi}{Kepler-17}
\newcommand{\koib}{Kepler-17b}

\newcommand{\hda}{HD~209458b}

\newcommand{\koicurCCkic}{KIC~10619192}
\newcommand{\koicurCCtwomass}{2MASS~19533486+4748540}

\newcommand{\koicurLCar}{\ensuremath{5.48\pm0.02}}			
\newcommand{\koicurLCrprstar}{\ensuremath{0.13031^{+0.00022}_{-0.00018}}}	
\newcommand{\koicurLCimp}{\ensuremath{0.268^{+0.014}_{-0.012}}}			
\newcommand{\koicurLCi}{\ensuremath{87\fdg{2}\pm0.15}}				
\newcommand{\koicurLCua}{\ensuremath{0.405\pm0.007}}			
\newcommand{\koicurLCub}{\ensuremath{0.262^{+0.013}_{-0.015}}}			

\newcommand{\koicurLCP}{\ensuremath{1.4857108\pm0.0000002}}	
\newcommand{\koicurLCPshort}{1.486}				
\newcommand{\koicurLCT}{\ensuremath{2455185.678035^{+0.000023}_{-0.000026}}}	

\newcommand{\koicurSMEteff}{\ensuremath{5630\pm 100}}	
\newcommand{\koicurSMEfeh}{\ensuremath{0.3\pm0.1}}	
\newcommand{\koicurSMElogg}{\ensuremath{4.42\pm0.015}}	
\newcommand{\koicurSMEvsin}{\ensuremath{4.7\pm1.0}}	

\newcommand{\koicurYYmlong}{\ensuremath{1.06\pm0.07}}	%
\newcommand{\koicurYYrlong}{\ensuremath{1.02\pm0.03}}%
\newcommand{\koicurYYlogg}{\ensuremath{4.43\pm0.02}}	%
\newcommand{\koicurYYage}{\ensuremath{3\pm1.6}}			%
\newcommand{\koicurYYspec}{G}

\newcommand{\koicurRVK}{\ensuremath{419.5^{+13.3}_{-15.6}}}			
\newcommand{\koicurRVmean}{\ensuremath{+0.40\pm0.10}}			

\newcommand{\koicurPPlogg}{\ensuremath{3.55\pm0.02}}	%
\newcommand{\koicurPParel}{\ensuremath{0.02591^{+0.00037}_{-0.00036}}}	
\newcommand{\koicurPPrho}{\ensuremath{1.35\pm0.08}}		%
\newcommand{\koicurPPm}{\ensuremath{2.45\pm0.11}}		%
\newcommand{\koicurPPmlong}{\ensuremath{2.45\pm0.11}}%
\newcommand{\koicurPPr}{\ensuremath{1.31\pm0.02}}		%
\newcommand{\koicurPPrlong}{\ensuremath{1.31\pm0.02}}	%
\newcommand{\koicurPPteq}{\ensuremath{1570\pm200}}			%

\shorttitle{Discovery and characterization of \koib}
\shortauthors{D{\'e}sert et al.}\def\simgr{\,\hbox{\hbox{$ > $}\kern -0.8em \lower 1.0ex\hbox{$\sim$}}\,}
\def\simle{\,\hbox{\hbox{$ < $}\kern -0.8em \lower 1.0ex\hbox{$\sim$}}\,}

\begin{document}

\title{The hot-Jupiter Kepler-17b: discovery,  obliquity from stroboscopic starspots, and atmospheric characterization}

\author{Jean-Michel D\'esert\altaffilmark{1},
David Charbonneau\altaffilmark{1},
Brice-Olivier Demory\altaffilmark{2},
Sarah Ballard\altaffilmark{1},
Joshua A. Carter\altaffilmark{1},
Jonathan J. Fortney\altaffilmark{3},
William D. Cochran\altaffilmark{4},
Michael Endl\altaffilmark{4},
Samuel N. Quinn\altaffilmark{1},
Howard T. Isaacson\altaffilmark{5},
Fran\c{c}ois Fressin\altaffilmark{1},
Lars A. Buchhave\altaffilmark{7},
David W. Latham\altaffilmark{1},
Heather A. Knutson\altaffilmark{5},
Stephen T. Bryson\altaffilmark{8},
Guillermo Torres\altaffilmark{1},
Jason F. Rowe\altaffilmark{8},
Natalie M. Batalha\altaffilmark{9},
William J. Borucki\altaffilmark{8},
Timothy M. Brown\altaffilmark{10},
Douglas A. Caldwell\altaffilmark{11},
Jessie L. Christiansen\altaffilmark{8},
Drake Deming\altaffilmark{6},
Daniel C., Fabrycky\altaffilmark{3},
Eric B. Ford\altaffilmark{12},
Ronald L. Gilliland\altaffilmark{13},
Micha\"el Gillon\altaffilmark{14},
Micha\"el R. Haas\altaffilmark{8},
Jon M. Jenkins\altaffilmark{11},
Karen Kinemuchi\altaffilmark{15},
David Koch\altaffilmark{8},
Jack J. Lissauer\altaffilmark{1},
Philip Lucas\altaffilmark{17},
Fergal Mullally\altaffilmark{11},
Phillip J. MacQueen\altaffilmark{4},
Geoffrey W. Marcy\altaffilmark{5}, 
Dimitar D. Sasselov\altaffilmark{1},
Sara Seager\altaffilmark{2},
Martin Still\altaffilmark{15},
Peter Tenenbaum\altaffilmark{11},
Kamal Uddin\altaffilmark{16},
and
Joshua N. Winn\altaffilmark{2}}

\altaffiltext{1}{Harvard-Smithsonian Center for Astrophysics, 60 Garden Street, Cambridge, MA 02138; jdesert@cfa.harvard.edu}
\altaffiltext{2}{Massachusetts Institute of Technology, Cambridge, MA 02159, USA}
\altaffiltext{3}{Department of Astronomy and Astrophysics, University of California, Santa Cruz, CA 95064, USA}
\altaffiltext{4}{University of Texas, Austin}
\altaffiltext{5}{Department of Astronomy, University of California, Berkeley, CA 94720-3411, USA} 
\altaffiltext{6}{Solar System Exploration Division, NASA Goddard Space Flight Center, Greenbelt, MD 20771, USA}
\altaffiltext{7}{Neils Bohr Institute, Copenhagen University}
\altaffiltext{8}{NASA Ames Research Center, Moffett Field, CA 94035}
\altaffiltext{9}{San Jose State University}
\altaffiltext{10}{Las Cumbres Observatory Global Telescope, Goleta, CA 93117}
\altaffiltext{11}{SETI Institute/NASA Ames Research Center, Moffett Field, CA 94035}
\altaffiltext{12}{University of Florida, Gainesville, FL 32611}
\altaffiltext{13}{Space Telescope Science Institute, Baltimore, MD 21218}
\altaffiltext{14}{Institut d'Astrophysique et de G\'eophysique,  Universit\'e de Li\`ege,  All\'ee du 6 Ao\^ut 17,  Bat.  B5C, 4000 Li\`ege, Belgium} 
\altaffiltext{15}{Bay Area Environmental Research Inst./NASA Ames Research Center, Moffett Field, CA 94035} 
\altaffiltext{16}{Orbital Sciences Corporation/NASA Ames Research Center, Moffett Field, CA 94035} 
\altaffiltext{17}{Centre for Astrophyiscs Research University of Hertfordshire} 


\begin{abstract}

This paper reports the discovery and characterization of the transiting hot giant exoplanet \koib. 
The planet has an orbital period of $\koicurLCPshort$ days, and radial velocity measurements from the {\it  Hobby-Eberly Telescope} show a Doppler signal of $\koicurRVK$ $\ms$.
From a transit-based estimate of the host star's mean density, combined with an estimate of the stellar effective temperature \teff=$\koicurSMEteff$ from high-resolution spectra, we infer a stellar host mass of \koicurYYmlong ~\msun\ and a stellar radius of \koicurYYrlong ~\rsun.
We estimate the planet mass and radius to be  $\mpl = \koicurPPmlong\,\mjup$ and $\rpl = \koicurPPrlong\,\rjup$.
The host star is active, with dark spots that are frequently occulted by the planet. 
The continuous monitoring of the star reveals a stellar rotation period of 11.89 days, 8 times the the planet's orbital period; this period ratio produces \emph{stroboscopic} effects on the occulted starspots.
The temporal pattern of these spot-crossing events shows that the planet's orbit is prograde and the star's obliquity is smaller than 15$^\circ$.
We detected planetary occultations of \koib\ with both the {\it Kepler} and \spitzer\ {\it Space Telescopes}. 
We use these observations to constrain the eccentricity, $e$, and find that it is consistent with a circular orbit ($e<0.011$).
The brightness temperatures of the planet the infrared bandpasses are T$_{\rm{3.6~\micron}}$=$1880\pm100$~K and T$_{\rm{4.5~\micron}}$=$1770\pm150$~K. 
We measure the optical geometric albedo $A_g$ in the \kepler\ bandpass and find $A_g = 0.10 \pm 0.02$.
The observations are best described by atmospheric models for which most of the incident energy is re-radiated away from the day side.

\end{abstract}

\keywords{ planetary systems --- stars: individual (\koib,
\koicurCCkic, \koicurCCtwomass) --- eclipses --- techniques: photometry }

\section{Introduction}\label{intro}

NASA's \kepler\ mission is a space-based photometric telescope dedicated to finding Earth-size planets in the habitable zones of their host stars and to determining the frequency and characteristics of planetary systems around Sun-like stars \citep{borucki10}.
While monitoring nearly continuously the brightness of about 150,000 dwarf stars since May 2009 to achieve the above goals, it accumulates an extraordinary volume of data allowing atmospheric studies of the transiting hot-Jupiters present in the field of view \citep{borucki09}. The first four months of observations were released and 1235 transiting planet candidates were reported \citep{borucki11}. Because of their short periods and relatively deep transit depths, several hot-Jupiter candidates were discovered early on during the first weeks of the mission. Amongst those, the {\it Kepler Object of Interest} KOI-203 was identified as a promising target, was selected for follow-up studies, and placed on the short-cadence target list for subsequent quarters.
This is the object of interest of the current study. 

The highly irradiated transiting hot-Jupiters currently provide the best opportunities for studying exoplanetary atmospheres in emission, during planetary occultations, when the exoplanets pass behind their parent stars \citep{seager00,sudarsky00,fortney05,barman05}. Light emitted from exoplanets was first detected from space at infrared wavelengths \citep{charbonneau05,deming05} and more recently in the optical \citep{alonso09a,snellen09,borucki09,alonso10} using the \corot\ and \kepler\ {\it Space Telescopes}.
Notably, \cite{rowe06,rowe08} used the {\it Microvariablity and Oscillations of Stars} ({\it MOST}) telescope to place a very stringent upper limit on the depth of the occultation of \hda.

The hot-Jupiters detected by \corot\ and \kepler\ are particularly good targets for studying planetary atmospheres, because the precise and nearly uninterrupted photometric surveillance provided by the space satellites can allow the planetary occultations to be detected at optical wavelengths \citep{snellen09,borucki09,desert11b}. 
Obtaining multiple wavelength observations of the relative depths of planetary occultations is necessary to constrain the broad band emergent spectra. Such observations are fundamental to understanding the energy budget of these objects \citep{sudarsky03,burrows05,burrows08,spiegel10} and for comparative exoplanetology.

We confirm here the planetary status of KOI-203, designating it as \koib, and study its atmosphere.
We first describe the \kepler\ observations and transit modeling as well as the follow-up observations used to confirm this planet, including an orbital solution using radial velocities obtained with the {\it High Resolution Spectrometer} (HRS) at the {\it Hobby-Eberly Telescope} (HET).
We then study the impact of the stellar variability on the transit lightcurves and we use occulted dark starspots to constrain the stellar obliquity.
We finally combine occultation measurements obtained in the optical with \kepler\ and in the infrared with \spitzer\ to learn about the atmospheric properties of \koib.

We first describe the observations, time series and analysis of the \kepler\ photometry in Sect.~2, then we describe the follow-up observations that confirm the planet detection in Sect.~3 and then discuss the stellar properties in Sect.~4. In Sect.~5, we describe the occultation measurements obtained from the visible and infrared. We present a global Monte-Carlo analysis of the complete sample of observations in Sect.~6 and finally discuss our findings in Sect.~7.

\section{\kepler\ observations}
\label{sec:kepler}

\begin{figure*}[h!]
\begin{center}
 \includegraphics[height=3in]{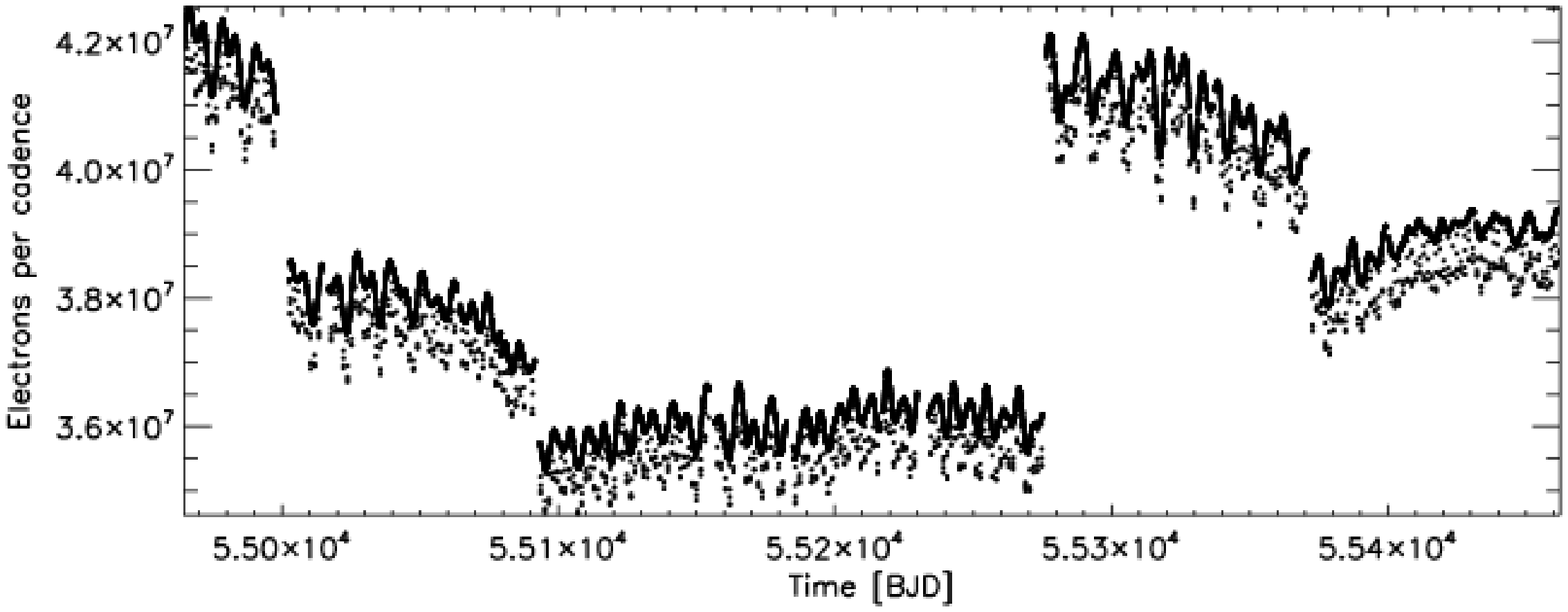} 
 \caption{Kepler raw long-cadence lightcurve of \koi, from Quarter 0 to Quarter 6. The drop of flux occurring every 1.48 days corresponds to the planetary transits. The stellar activity shows $3\%$ flux variation and with a period of 11.89 days.}
  \label{fig:rawtlc}
\end{center}
\end{figure*}

Observations of the \kepler\ field commenced in May 2009 with Quarter 0 (Q0);
the data that we describe here are the \kepler\ science data of \koi\ from Quarter~0 to 6 (Q0-Q6). 
The \kepler\ observations were gathered almost continuously during 16.7 months.
These observations have been reduced and detrended by the \kepler\ pipeline \citep{jenkins10a}. 
The \kepler\ bandpass spans $423$ to $897$~nm for which the response is greater than 5\% \citep{batalha10,bryson10}. 
This wavelength domain is roughly equivalent to the V+R-band \citep{koch10a}.
The target \koi\ was identified in the \emph{Kepler Input Catalog} \citep{brown11} (\koicurCCtwomass, KIC 10619192, r= 14.08 mag).
Because of the transit like events, the object was then considered for follow-up studies, identified as {\it Kepler Object of Interest} KOI-203. 
The pipeline produces both calibrated light curves (PA data) for individual analysis and corrected light curves (PDC) which are used to search for transits.
This paper presents results which are measured from PA data only.
They consist of long cadence integration time (29.426~minutes) for Quarters~0 and 1 \citep{caldwell10,jenkins10b} and long and short cadence (1~minute) for Quarters~2 to 6 \citep{gilliland10}. 
The pipeline provides time series with times in barycentric corrected Julian days, and flux in photo-electrons/per cadence.
The raw nearly continuous photometry of \koi\ is presented in Figure~\ref{fig:rawtlc}.
We measure the transit parameters from the \kepler\ observations as described below (see Section~\ref{sec:mcmc}).
We present the normalized, phase-folded and combined transit light curve obtained at short cadence in Figure~\ref{fig:keplerlc} from which we measure the transit parameters.

\section{Follow-up observations}

\subsection{Reconnaissance spectroscopy}
\label{sec:recon}

As described in detail by \citet{gautier10}, the follow-up observations of {\em Kepler} planet candidates involve
reconnaissance spectroscopy to look for evidence of astrophysical false positives responsible for the observed transits. These false positives include single- and double-lined binaries, certain types
of hierarchical triples and even some background eclipsing binaries, which would show velocity variations and/or composite spectra that are readily detectable by the modest facilities used for these
reconnaissance observations. 
As described below, we also use these spectra to estimate the effective temperature, surface gravity, metallicity, and rotational and radial velocities of the host star.

On 25 Apr 2011 UT, we obtained a spectrum of \koi\ using the fiber-fed  {\it Tillinghast Reflector Echelle Spectrograph} (TRES; \cite{furesz08}) on the 1.5m  {\it Tillinghast Reflector} at the  {\it Fred Lawrence Whipple Observatory}, on Mt. Hopkins, AZ.  
The spectrum was taken with the medium fiber, which has a resolving power of $\lambda/\Delta\lambda$=44,000 and a wavelength coverage of about 3850 to 9100 Angstroms.  
The exposure time was 80 minutes.
The spectrum was extracted and rectified to intensity vs. wavelength using standard procedures described by \cite{buchhave10}.  
The extracted spectrum has a signal-to-noise ratio of 17.5 per resolution element.

We performed cross-correlations against a grid of synthetic stellar spectra, calculated by John Laird for a grid of Kurucz model atmospheres  \citep{kurucz79}, using a line list developed by Jon Morse.
The grid is coarse -- 250 K in effective temperature, \teff; 0.5 in log
surface gravity, \logg; 0.5 in log of the metallicity compared to the
sun, \mh; and 2 km s-1 in rotational velocity, \vrot\ -- so rather than simply adopting the parameters from the template with the best correlation coefficient, we fit a surface to the correlation peak heights to arrive at a refined classification.  
However, given the degeneracies between \teff, \logg, and \mh, the quality of the spectrum is not sufficient to determine all three at once.  
As such, we have fixed $\logg=\koicurSMElogg$ as inferred from the best-fit of the Kepler photometry.  
The analysis yields \teff=$\koicurSMEteff$~K, \mh=$\koicurSMEfeh$, $\vsini=\koicurSMEvsin$~\kms. 
When corrected for the orbital motion of \koi\ and the TRES zero point offset, determined by long-term monitoring of the IAU RV standard HD~182488, we find the absolute mean systemic velocity of \koi\ to be $-23.82\pm0.10$~\kms. 
Note that this does not include any uncertainty in the absolute velocity of HD~182488, which we take to be -21.508~\kms, as observed by \cite{nidever02}.

We also obtained Keck HIRES spectra and estimated the line strengths of \sval=$0.322\pm0.01$\ and \logrhk=$-4.61$ for \koi\ (assuming B-V=0.82).
The \caii\ line strengths are a good indicator of the stellar activity \citep{isaacson10}.

\subsection{Imaging}

The astrometry derived from the {\em Kepler} images themselves, when combined
with high-resolution images of the target neighborhood, provides
a very powerful tool for identifying background eclipsing binaries
blended with and contaminating the target images \citep{batalha10}.  The astrometry of \koi\ indicated no significant offset during transits in any quarter, and computed offsets are well within the formal 3-sigma radius of confusion. Therefore \koi\ is considered to be the source for the transits observed in the \kepler\ lightcurves.

\begin{figure*}[h!]
\begin{center}
 \includegraphics[width=2in, angle=-90]{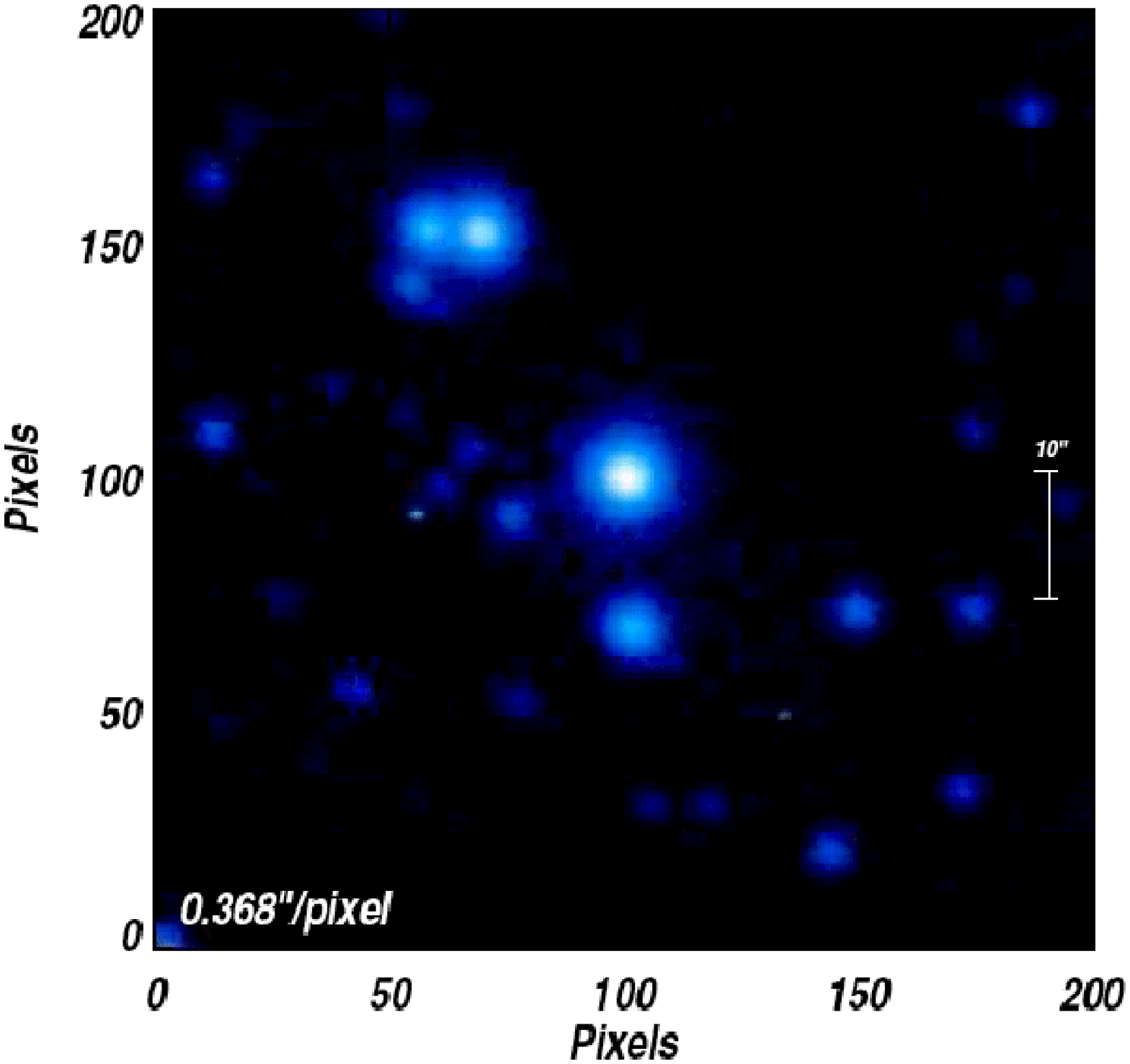} 
 \includegraphics[width=2in, angle=-90]{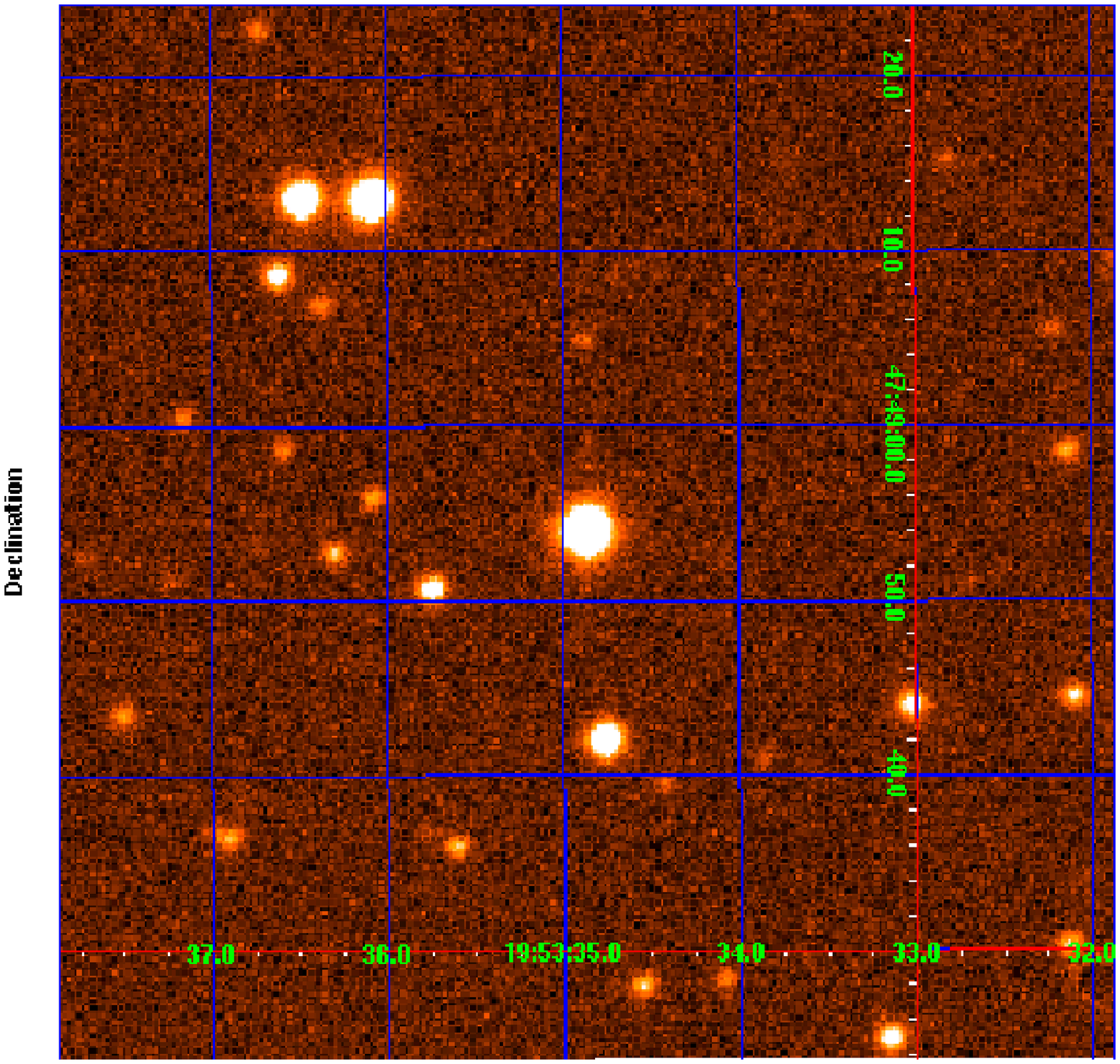} 
  \caption{The left panel shows a I-band natural seeing image of \koi\ taken with the Lick 1-meter telescope (1.2'x1.2'). The right panel shows a J-band image
of \koi\ taken with the wide field camera (WFCAM, 1'x1') on the United Kingdom Infrared Telescope
(UKIRT). These images confirm that there are no companions from 2" to 5" from the star's center, down to a limit of 19th magnitude.}
   \label{fig:image}
\end{center}
\end{figure*}

We obtained an I-band image at the Lick Observatory 1-meter, Nickel telescope with the Direct Imaging Camera (see Figure~\ref{fig:image}). The 1.0" arc-second seeing revealed no companions from 2" to 5" from the star's center, down to a limit of 19th magnitude. Similar conclusion were reached using UKIRT J-band images (Figure~\ref{fig:image}).

\subsection{Radial velocity}
\label{sec:RVs}

We obtained precise radial velocity (RV) follow-up observations of \koi\ with the HET (Ramsey et al.~1998) and its HRS spectrograph \citep{tull98} at McDonald Observatory. \koi\ was observed ten times in the 2010 observing season, from 2010 August 22 until 2010 November 22. 
The instrumental setup and observing mode are described in more detail in \cite{endl11}. 
\koib\ is the second planet confirmed with HET after Kepler-15b \citep{endl11}. 
We employed a ``snap shot'' strategy, using relatively short exposures of 1200 seconds, that yield a SNR sufficient to detect the radial-velocity signal of a hot-Jupiter. 
Thirteen spectra were taken with the I$_2$-cell in the light path to compute precise differential RVs. 
These spectra have a typical S/N-ratio of $32$ per resolution element. 
The radial velocity data are listed in Table~\ref{tab:RVs}.

\begin{figure*}[h!]
\begin{center}
 \includegraphics[width=5.5in, angle=-90]{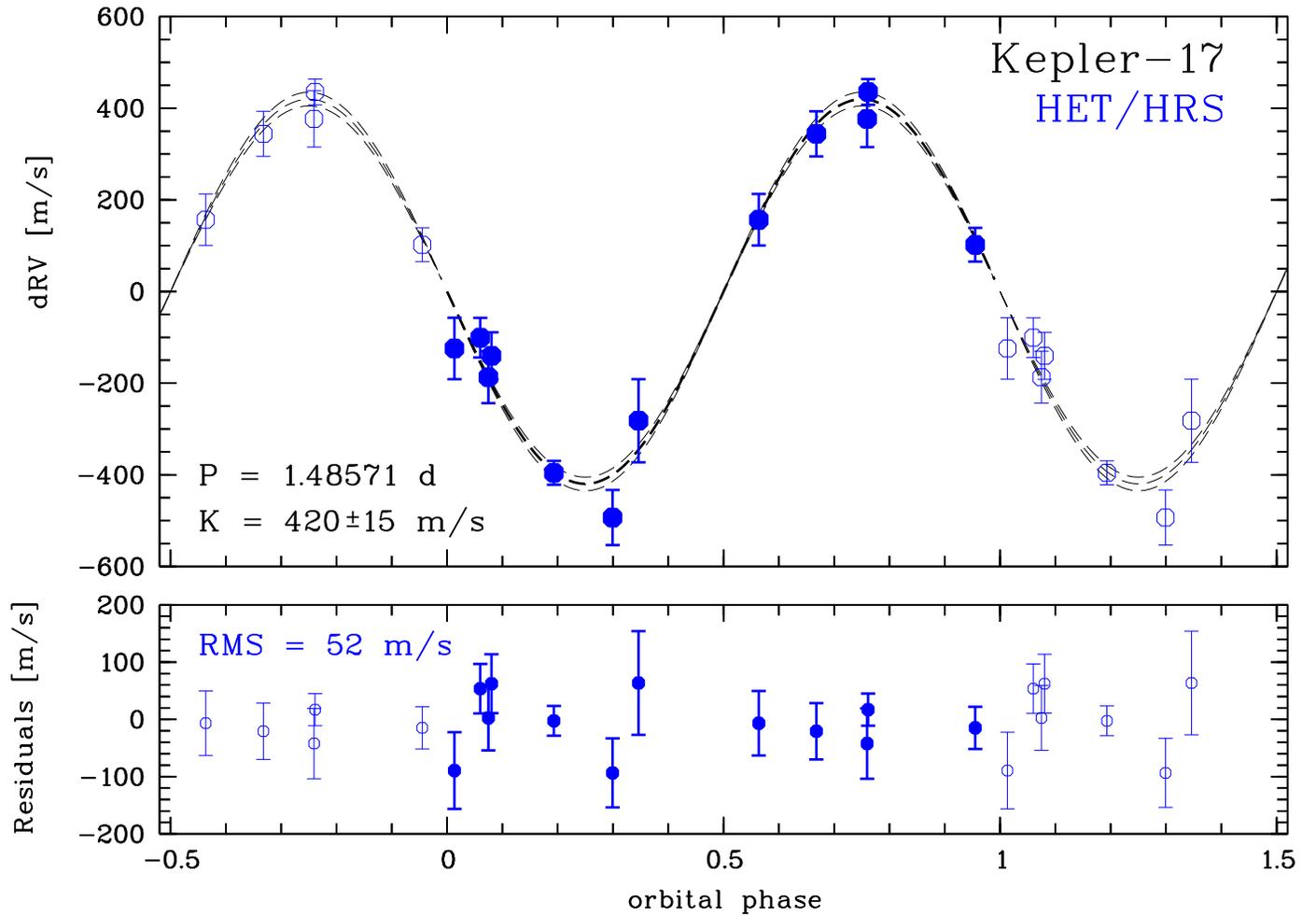}
 
  \caption{ Orbital solution for \koi. Top panel: the observed radial velocities
obtained with HET HRS are plotted
together with the velocity curve for a circular orbit with the period
and time of transit fixed by the photometric ephemeris. The thick dashed line corresponds to the best fit and the thin dashed lines correspond to models with $\pm 1\sigma$ error on the parameter K.  The $\gamma$
velocity has been subtracted from the relative velocities here and in
Table~\ref{tab:RVs}, and thus the center-of-mass velocity for the orbital solution is 0 by definition.
Bottom panel: The velocity residuals from the orbital solution.  The rms of the velocity residuals is 52\,\ms.}
   \label{fig:RVs}
\end{center}
\end{figure*}

We use {\it Gaussfit}, the generalized least-squares software of \citet{jefferys88}  to 
fit a Keplerian orbit to the HRS radial velocity data. 
Only the velocity zero-point and the radial velocity semi-amplitude $K$ are included
as free parameters in the fitting process. 
We first fitted the radial-velocity data alone, requiring the orbit to be circular ($e=0$) and adopting the ephemeris derived from the \kepler\ photometry. 
The best-fit orbit has a $K$ of $420\pm15$\,m\,s$^{-1}$, a $\chi^{2}_{\rm red}$ of $0.9$ 
and a residual rms scatter around the fit of 52~\,m\,s$^{-1}$. 
The radial velocity data and the orbital solution are shown in Figure~\ref{fig:RVs}.

We determined the spectral line bisectors, which are a measure of line asymmetry, from the HET spectra to test if the radial velocity variations could be caused by distortions in the spectral line profiles due to contamination from a nearby unresolved eclipsing binary.
We can only use a small fraction of the available spectral range that is not contaminated by the iodine absorption cell (5000--6400 Angstrom) and thus the uncertainties in the bisector velocity span (BVS) are quite large with an average uncertainty of 99~\,m\,s$^{-1}$. 
The RMS of the bisector measurements is 146~\,m\,s$^{-1}$. 
There is no evidence of a correlation between the velocities and the bisectors, which supports the interpretation that the velocity variations are due to a planetary companion (e.g. \citealt{queloz01})

\section{Stellar Parameters}
\label{sec:stellar}

We derive the mass, radius, and age of the host star using the method described by \cite{torres08}. 
We first created a set of stellar evolution models from the Yonsei-Yale (Y$^{2}$) series by \cite{yi01}, with corrections from \cite{demarque04}. We employed their interpolation software which accepts as inputs the age of the star, the iron abundance, and the abundance of $\alpha$-elements (relative to solar) for which we assume the solar value, and outputs a grid of stellar isochrones corresponding to a range of masses. We evaluated a set of isochrones at ages from  0.1 to 14 Gyr (at intervals of 0.1 Gyr) and stellar metallicities spanning a range of $3\sigma$ (at intervals of 0.01 dex) from the best-fit metallicity derived from spectra of \feh=$\koicurSMEfeh$. We then performed a spline interpolation of each output table at a resolution of 0.005 in effective temperature \teff, the log of the surface gravity log$(g)$, and the stellar luminosity $L_{\star}$. We evaluated the physical radius corresponding to each stellar model via log$(g)$ and the mass of the star, though it is also possible to convert to physical radius using the model stellar luminosity and effective temperature (assuming $L_{\star}=4\pi R_{\star}^{2}\sigma T^{4}$); in practice these conversions give identical results.

We fitted for the stellar mass and radius using Newton's version of Kepler's third law in the manner employed by \cite{seager03}, \cite{sozzetti07} and \cite{torres08}.
We assumed that the planetary mass is negligible when compared to the mass of the host star.
Using the Markov Chain Monte Carlo sequence of $a/R_{\star}$, and generating a series of Gaussian random realizations of \feh\ and \teff\ using the values and error bars derived from spectroscopy of \teff=$\koicurSMEteff$~K and \feh=$\koicurSMEfeh$, respectively, we located the best isochrone fit at each realization using the $\chi^{2}$ goodness-of-fit given in Equation \ref{eq:chisquared}. 

\begin{equation}
\chi^{2}=\left(\frac{\Delta a/R_{\star}}{\sigma_{a/R_{\star}}}\right)^{2}+\left(\frac{\Delta \teff}{\sigma_{\teff}}\right)^{2}+\left(\frac{\Delta[\mbox{Fe/H}]}{\sigma_{\mbox{[Fe/H]}}}\right)^{2},
\label{eq:chisquared}
\end{equation}

Using the output of the MCMC chain of $a/R_{\star}$ ensures that any correlations between parameters, which are reflected in the chain, are properly incorporated into our estimate of the stellar parameters.

\begin{figure*}[h!]
\begin{center}
\includegraphics[height=5in]{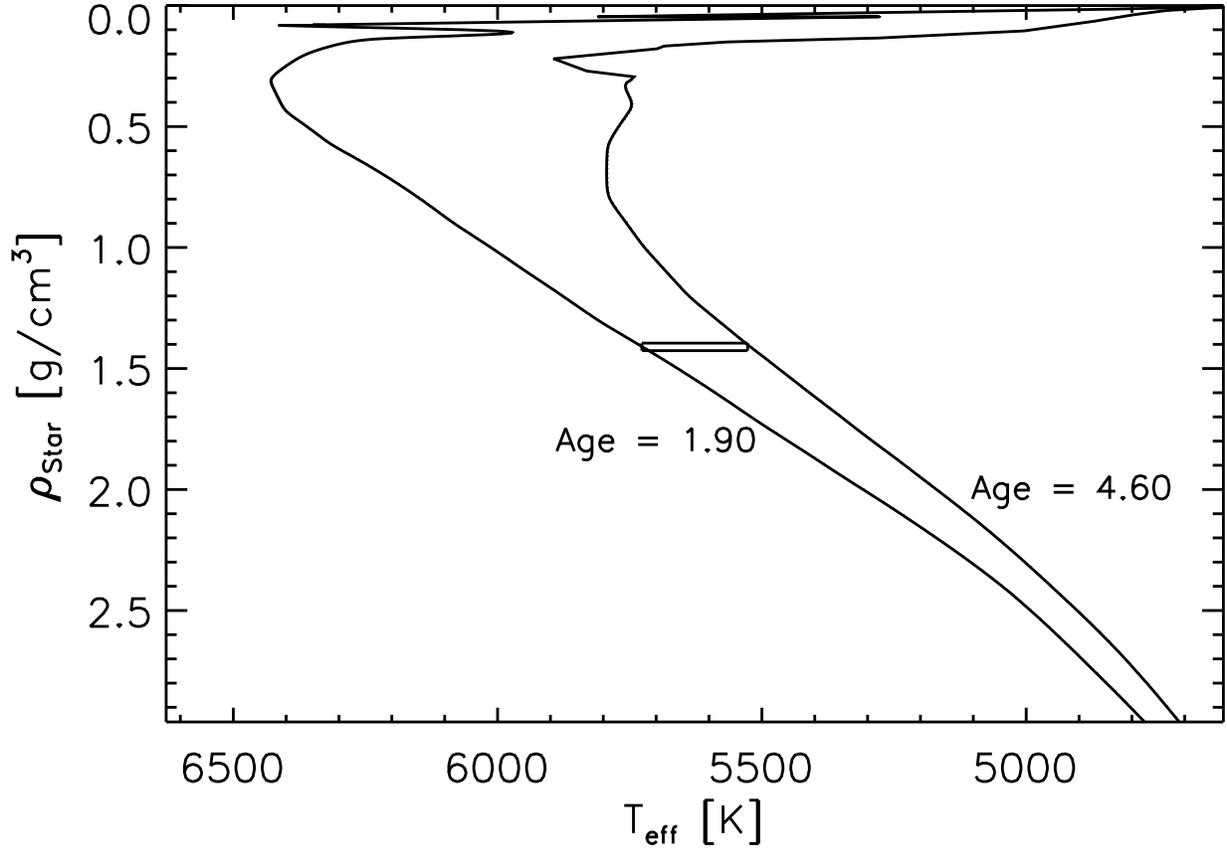} 
 \caption{Stellar isochrones in the range of the observed \teff\ and \rhostar. The small rectangle corresponds to the range of possible solutions.}
  \label{fig:isochrone}
\end{center}
\end{figure*}

We then assign a weight to the likelihood of each stellar model in the chain, applying a prior for the initial mass function (IMF) which assumes a Salpeter index. The number of stars of each mass and age, per 1000 stars, is generated by the interpolation software provided by \cite{yi01} for several IMF assumptions, including the Salpeter IMF. We designate the weight assigned to each stellar model in the chain by normalizing to the highest IMF value within the sample; in practice, the weights vary from 0.3 to 1 (from the least to most likely). The IMF prior changes the final answer by less than half a $\sigma$  for all parameters. We then incorporate this likelihood by discarding members of the chain according to their weight, where the weight is equal to the likelihood of remaining sampling; about 40\% of the original chain remains intact after this stage. The value for each stellar parameter is then assigned from the median of this weighted distribution, with the error bars assigned from the nearest $\pm34\%$ of values. In this way, we find $M_{\star}$= $1.061^{+0.045}_{-0.040}$ $M_{\odot}$, $R_{\star}$= 1.019 $\pm$ 0.014 $R_{\odot}$, and an age = $2.9^{+1.5}_{-1.6}$ Gyr. These uncertainties are statistical as they exclude possible systematic uncertainties in the stellar models.
The best-fit solution is presented in Figure~\ref{fig:isochrone}.

We caution that stellar isochrones are only poorly constrained for this faint star.
The estimated error bars on the stellar parameters are smaller than what is expected for such a star of this magnitude.
This is because we set the stellar gravity as a fixed value in our analysis, since the stellar spectrum has a low signal-to-noise.
We set the stellar gravity to the value we derive from the stellar density measured using the \kepler\ photometry (see Section~\ref{sec:recon}).
The quality of the combined transit lightcurve allows us to measure the stellar density with a high precision that tightly constrains the isochrone fits (see Figure~\ref{fig:isochrone}).
However, a recent study of the physical properties of the stellar components of the transiting exoplanets by \citet{southworth11} shows that the current stellar models are not determined at better than 1\% in terms of radius and 2\% in mass (median values); more conservatives values are 3 and 5\% for the stellar radius and mass respectively.
We adopt these conservatives values and propagate the systematic errors arising from the dependence on stellar theory to the final errors.
Adding the statistical and systematical error quadratically, we find $M_{\star}$= $1.061\pm0.07$ $M_{\odot}$, $R_{\star}$= 1.019 $\pm$ 0.03 $R_{\odot}$.
This study shows that \koi\ is consistent with a main-sequence $\koicurYYspec$ dwarf star, its radius and \teff\ are indistinguishable from Solar.

\section{Occultations of \koib\ }
\label{sec:eclipse}

\subsection{Occultations from the \kepler\ photometry}
\label{sec:eclipsekepler}

The first method we use to measure the occultations from the \kepler\ observations is described by \citet{desert11b}.
We search for the occultation events in the short cadence light curves. 
We normalize and combine 173 occultations events.
The best-fit and the maximum depth as a function of the orbital phase are found very close to the orbital phase of $0.5$ as expected for a circular orbit. 
We estimate the significance of this detection by measuring the occultation depth, ephemeris and associated errors using a bootstrap Monte-Carlo analysis. 
We find that the planet has an occultation depth of $58\pm10$~ppm.

\begin{figure*}[h!]
\begin{center}
 \includegraphics[width=5.5in]{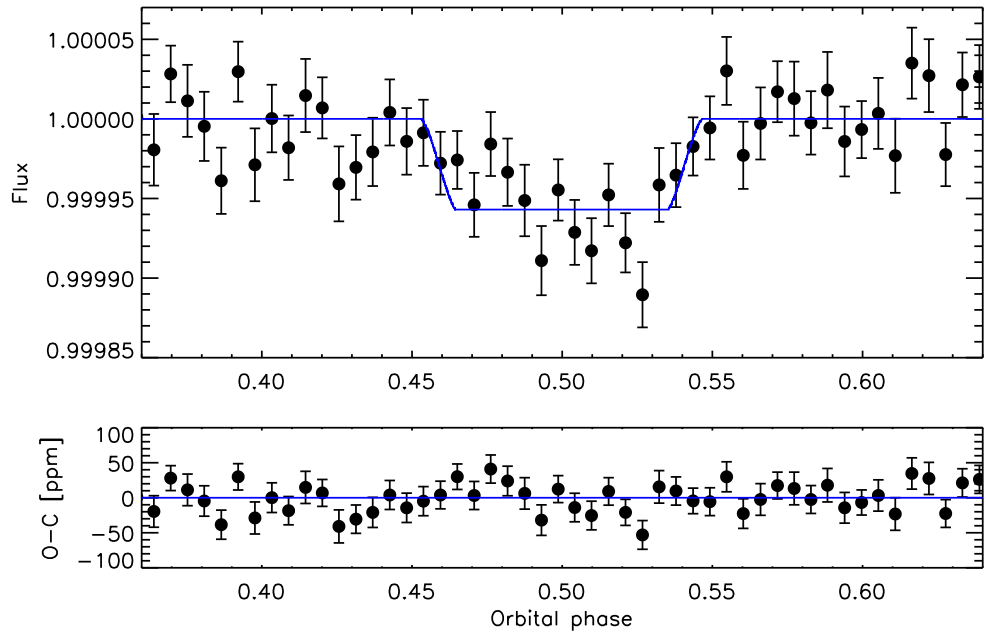}
  \caption{Phase-folded and normalized \kepler\ occultation lightcurve of \koib\ binned by 15~minutes (top panel). The best-fit model of the occultation in the \kepler\ bandpass is overplotted in blue and the residual from this best-fit are shown in the bottom panel.}
   \label{fig:keplerecllc}
\end{center}
\end{figure*}

In a second method, we include the occultations in a Markov Chain Monte Carlo global analysis as described below in the Section~\ref{sec:mcmc}.
We present the normalized, folded, combined and binned per 15 min light curve obtained at long cadence in Figure~\ref{fig:keplerecllc} from which we measure the occultation depth with the MCMC.

\subsection{\spitzer\ observations and photometry}
\label{sec:spitzer}

\begin{figure*}[h!]
\begin{center}
 \includegraphics[width=5.5in]{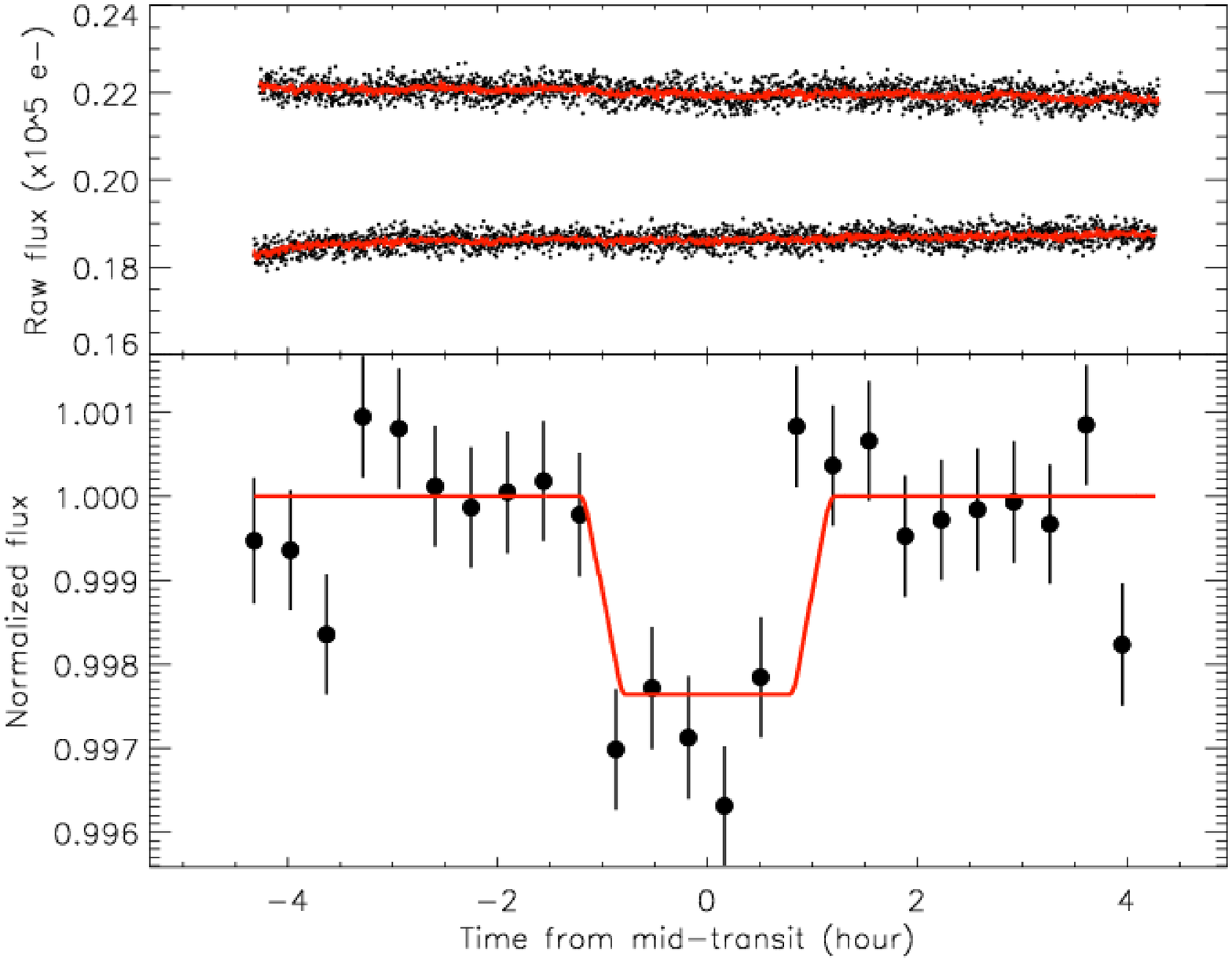}
 \includegraphics[width=5.5in]{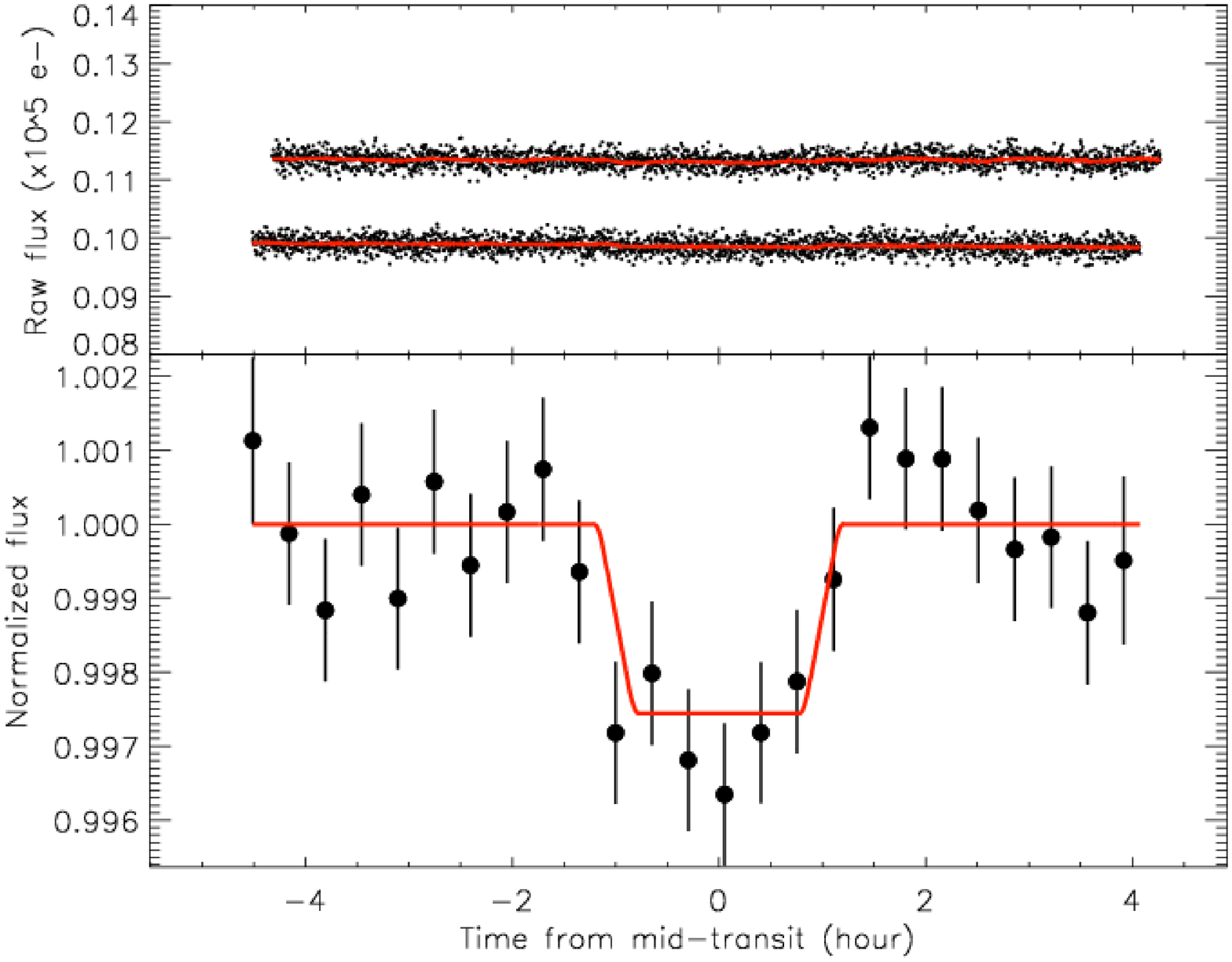}
 \caption{\spitzer\ occultation lightcurves of \koi\ observed in the IRAC band-pass at 3.6 (top) and 4.5~\micron\ (bottom). Two visits were obtained for each \spitzer\ bandpass. Top panels~: raw and unbinned lightcurves. The red solid lines correspond to the best-fit models which include the time and position instrumental decorrelations as well as the model for the planetary occultations (see details in Sect.~\ref{sec:spitzer}). Bottom panels~: corrected and normalized occultation lightcurves with their best-fit models (in red). The data are binned in 25~minutes intervals (50 data points per bin).}
   \label{fig:spitzerlc}
\end{center}
\end{figure*}

The method we use to measure the occultations from the \spitzer\ observations is described in \citet{desert09}.
\koi\ was observed during four occultations between August 2010 and January 2011 with \wspitzer/IRAC \citep{werner04,fazio04} at 3.6 and 4.5~\micron\ (program ID 60028). Two occultations were gathered per bandpass and each visit lasted approximately 8.6~h. 
The data were obtained in full-frame mode ($256\times256$ pixels) with an exposure time of
30.0~s per image and yielded 2461 images per visit. 
The method we use to produce photometric time series from the images is described in \cite{desert11a}.
It consists of finding the centroid position of the stellar point spread function (PSF) and performing aperture photometry using a circular aperture on individual Basic Calibrated Data (BCD) images delivered by the \emph{Spitzer} archive.
The optimal apertures were found to be $2.5$~pixels in radius. 
The final photometric measurements used are presented in Table~\ref{tab:spitzer}. 
The raw time series are presented in the top panels of Figure~\ref{fig:spitzerlc}.
Telescope pointing drift results in fluctuations of the stellar centroid position, which, in
combination with intra-pixel sensitivity variations, produces
systematic noise in the raw light curves (upper panel Figure~\ref{fig:spitzerlc}).  
A description of this effect, known as the pixel-phase effect, is given in the \spitzer/IRAC data handbook \citep{reach06} and is well known in exoplanetary studies (e.g. \citealt{charbonneau05,knutson08}). 
To correct the light curve, we define a baseline function that is the
sum of a linear function of time and a quadratic function (with four parameters) of the X and Y centroid positions. 
We find that the point-to-point scatter in the photometry gives a typical signal-to-noise ratio ($S/N$) $90$ and $110$ per image at 3.6 and 4.5~\micron\ respectively. 
These correspond to 85\% of the theoretical signal-to-noise.

We simultaneously fit the instrumental functions with all the parameters, measured the occultation depths for each individual visit, and report report the values in Table~\ref{tab:spitzer}.
The measurements per bandpass agree at the 1-$\sigma$ level.
The weighted mean of the transit depths are $0.250\pm0.030$\% and  $0.310\pm0.035$\%  at 3.6 and 4.5~\micron\ respectively. 

We also included the \spitzer\ observations in the global MCMC analysis described below.

\section{Markov Chain Monte Carlo analysis}
\label{sec:mcmc}

The analysis of the \kepler\ photometry and the determination of the
stellar and planetary parameters for \koi\ follows procedures similar to those reported in \citet{borucki10}.
We check and confirm these results by an independent analysis with the global Markov Chain Monte-Carlo (MCMC) algorithm presented in \citet{gillon09, gillon10}. 
This MCMC implementation uses the Metropolis-Hasting algorithm to perform the sampling.
Input data to the MCMC include the \kepler\ and \spitzer\ photometric measurements and the radial-velocity data.

\textit{Kepler} Short Cadence (SC) data allow a precise determination of the transit parameters and more specifically allow a fit for the limb-darkening (LD) coefficients. We therefore assumed a quadratic law and used $c_1 = 2 u_1 + u_2$ and $c_2 = u_1 - 2 u_2$ as jump parameters, where $u_1$ and $u_2$ are the quadratic coefficients. Those linear combinations help in minimizing correlations on the uncertainties of $u_1$ and $u_2$ \citep{holman06}.

The MCMC has the following set of 13 jump parameters: the planet-to-star flux ratios in the \kepler\ and \spitzer\ bandpasses, the transit depth, the impact parameter $b$, the transit duration from first to fourth contact, the time of minimum light, the orbital period, $K' = K \sqrt{1-e^2} P^{1/3}$, where $K$ is the radial-velocity semi-amplitude, the two LD combinations $c_1$ and $c_2$ and the two parameters $\sqrt{e}\cos \omega$ and $\sqrt{e}\sin \omega$ \citep{Anderson2011}. 
A uniform prior distribution is assumed for all jump parameters.
Baseline model coefficients are determined for each lightcurve with the SVD method \citep{press92} at each step of the MCMC. 
Correlated noise is accounted for following \citet{winn08} and \citet{gillon10}, to ensure reliable error bars on the fitted parameters. For this purpose, we compute a scaling factor based on the standard deviation of the binned residuals for each lightcurve with different time bins. The error bars are then multiplied by this scaling factor. We obtained a mean scaling factor of 1.19 for all photometry. 
The results from the MCMC analysis are presented in Table~\ref{tab:param}


\section{Results and Discussion}\label{results}

\subsection{Characteristics of the system \koi\ }
\label{sec:system}

Adopting the stellar mass and radius for \koi\ that we found in Section~\ref{sec:stellar}, we obtain a mass of $\mpl = \koicurPPmlong\,\mjup$ and a radius of $\rpl = \koicurPPrlong\,\rjup$ for the planet, 
which leads to a density of $\rhopl = \koicurPPrho\,\gcmc$, and a surface gravity $\loggpl = \koicurPPlogg $ (cgs). We note that the surface gravity can be derived from the photometry and the velocimetry only \citep{southworth07}.
The position of \koib\ on the mass/radius diagram is fairly common and it appears to be slightly inflated compared to models that include the effects of stellar irradiation (e.g. \citealt{latham10}).

The current upper limit on the orbital eccentricity $e$ from radial velocity measurements is consistent with zero for \koib.
We also measure the mid-occultation timing offset from both \kepler\ and \spitzer\ observations. 
The determination of the timing of the secondary eclipse constrains the planet's orbital eccentricity. 
Our estimate for the best-fit timing offset translates to a constraint on $e$ and the argument of pericenter $\omega$. 
The timing is used to constrain the $e\cos(\omega)$ and we find that $e < 0.011$ at the $1-\sigma$ level.
This upper limit implies that the orbits of these objects are nearly circular unless the line of sight is aligned with the planet's major orbital axis, i.e. the argument of periapse $\omega$ is close to 90\degr~or 270\degr.

\begin{figure*}[h!]
\begin{center}
 \includegraphics[height=5in]{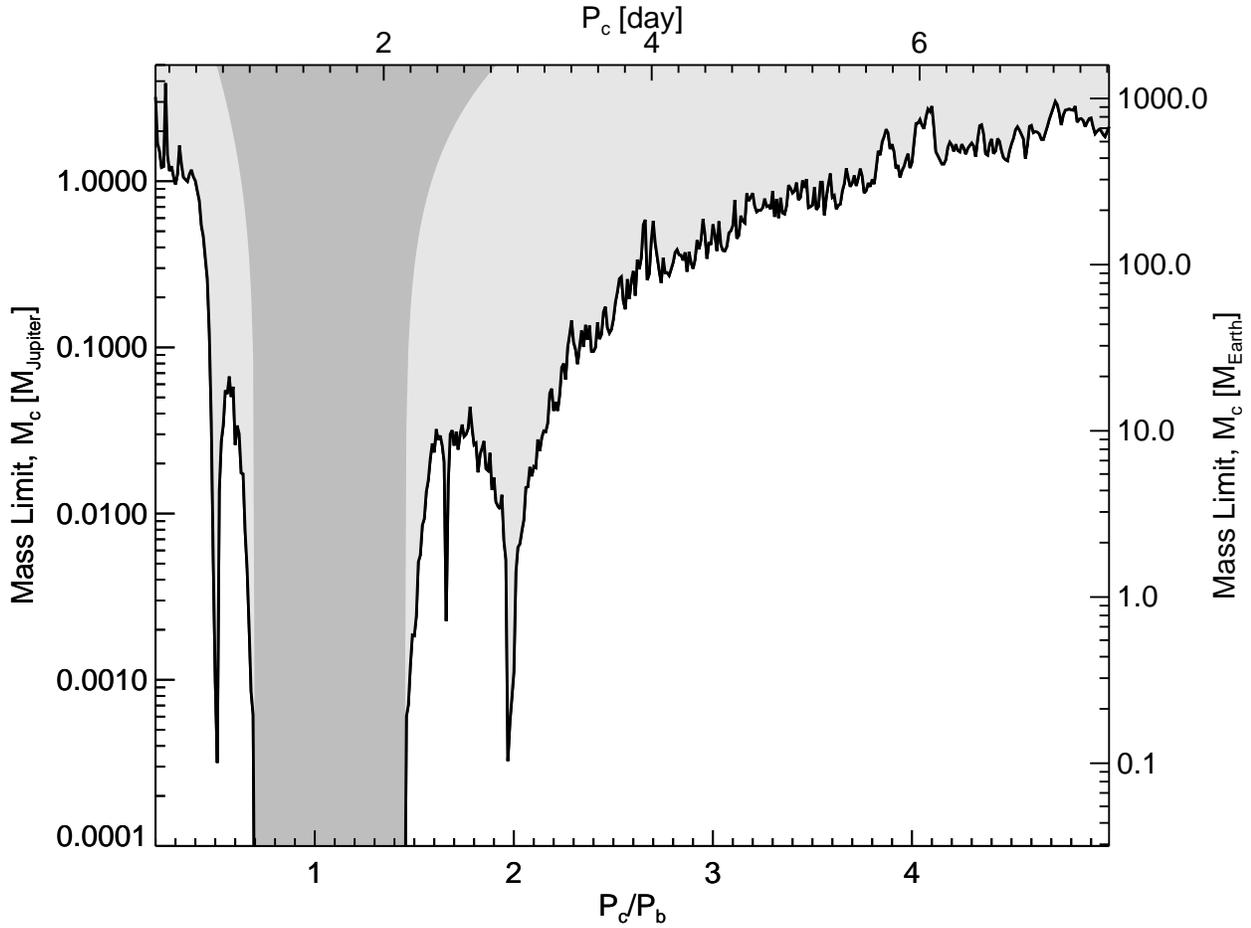} 
 \caption{Upper mass limits for a hypothetical perturber as a function of perturber period normalized to the period of \koib, $P_{c}/P_{b}$. The dark gray zone corresponds to the region of dynamical instability, the white zone corresponds to the region where the presence of a companion with a minimum mass would be permitted by the current limits on TTVs.}
  \label{fig:ttvs}
\end{center}
\end{figure*}

We fit a linear function of transit epoch and find $T_{c}(0)=\koicurLCT$ (BJD$_{utc}$) with a period $P=\koicurLCP$ days.
We also fit each mid-transit time individually and compare each one to the expected linear ephemeris to obtain a Observed-minus-Computed (O-C) time series. 
The O-C times shows a scatter that is most probably caused by stellar spot-induced shifts, since the star is active (see Section~\ref{starvar}).
Formally, the individual fits are consistent with the linear model, therefore we do not consider this to be a significant detection of timing anomalies.
Since there is no clear evidence for transit timing variations (TTVs; \citealt{agol05,holman05}), we use the timing data to place upper limits on the mass of a hypothetical second planet that would perturb the orbit of the transiting planet using the procedure described in \cite{carter11}.
Figure~\ref{fig:ttvs} shows the constraints on the perturber mass as a function of period ratio, as determined from this analysis.
The mass constraints on the perturber are more restrictive near the mean-motion resonances and most restrictive at the low-order resonances, particularly for the interior and exterior 2:1 resonances. 
For example, a perturber at the interior 2:1 resonance having a mass near that of Mars would have induced TTVs detectable in the present data.

\subsection{Stellar variability of \koi\ }
\label{starvar}
\begin{figure*}[h!]
\begin{center}
 \includegraphics[height=3in]{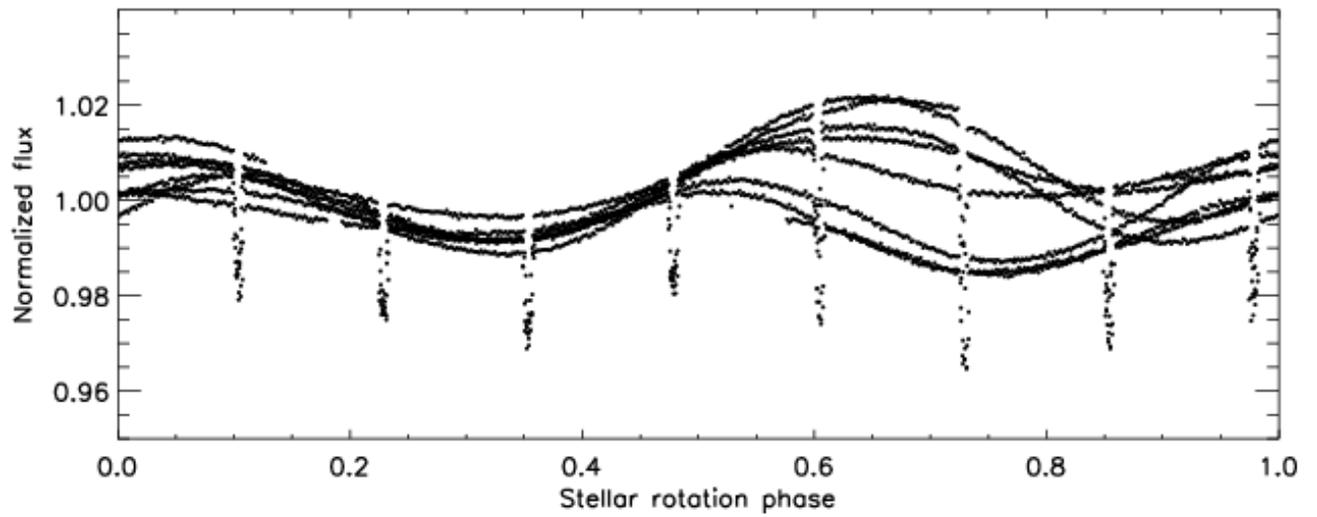} 
 \caption{Stellar phase folded Kepler normalized long-cadence lightcurves of \koi\ from Quarter 3.  
The flux is normalized to its median value measured over the whole quarter.
The stellar rotation period is 8 times the planet's orbital period. 
This integer period ratio allows one to see the transits at the same stellar phase modulo 8. 
The stellar variability exhibits $3\%$ flux variation with a period of approximately 11.9 days.}
\label{fig:stellarcycle}
\end{center}
\end{figure*}

\begin{figure*}[h!]
\begin{center}
 \includegraphics[height=5in]{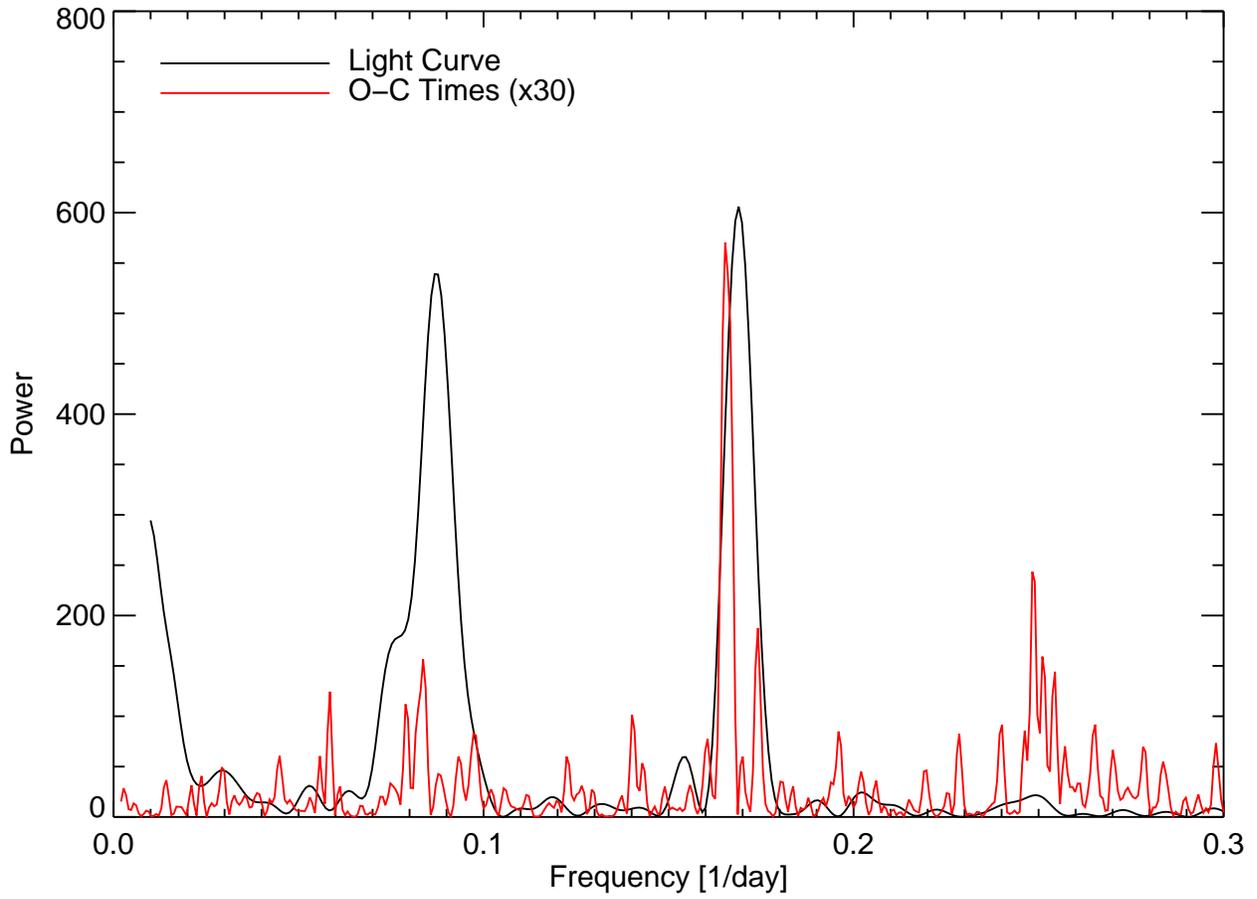} 
 \caption{Lomb-Scargle periodograms of the \kepler\ photometry (in black) and of the observed minus computed (O-C) mid-transit times (in red) magnified by 30 for comparison purposes. The periodogram from the photometry shows two main peaks at 5.95 and 11.9 days while the O-C periodogram exhibits a dominant peak at 6 days.
}

\label{fig:ls}
\end{center}
\end{figure*}

The \kepler\ photometry exhibits a quasi periodic flux modulation of about $3\%$ with a period of  11.9~days (Figure~\ref{fig:rawtlc}). 
This period corresponds to 8 times the planet's orbital period (Figure~\ref{fig:stellarcycle}). 
The stellar variability can be quantified from a Lomb-Scargle periodogram that reveals two peaks at 5.95 and 11.9 days (see Figure~\ref{fig:ls}). 
The position of the peak on the periodogram and its width provide a measurement of the stellar rotation period and its error, respectively. We measured a stelllar rotation period of $11.9\pm1.1$~days.
The presence of active stellar spots localized on opposite hemispheres can best explain the two peaks seen in the periodogram. 
We find that the ratio between the planet's orbital period and the stellar rotation period is  $8.0\pm0.7$~days.
We note that this intriguing integer ratio of 8 could potentially reveal the signature of stellar-planet interactions.

\subsubsection{Using occulted starspots to infer the orbital obliquity}
\label{starspots}

The lightcurve of \koi\ shows substantial deformation of the planet's transit profiles which we interpret as occultations of dark starspots (Figure~\ref{fig:keplerlc}).
We co-added the transit lightcurves that occurred at epochs modulo the stellar rotation period (modulo 8 planetary orbital periods).
We present the resulting 8 transit lightcurves in Figure~\ref{fig:spotseq}; each contains about 22 individual transit lightcurves. 
We identified that the planet crosses spots of identical shapes every 8 transits producing a \emph{stroboscopic} effect. 
Therefore, we conclude that starspots come back to the same position on the transit chord after 8 orbits or 11.89 days (one stellar rotation period).
Because of this stroboscopic effect, the scatter measured during the spot-crossing phase of each of these lightcurve residuals is similar to the scatter of the residuals measured outside the transit. 
This further supports the assertion that the same spots are crossed by the planet every 8 transits, at identical orbital phases, because the residuals measured from a transit light curve obtained from a planet occulting a random distribution of spots are expected to be larger during the transit event compared to the out-of-transit monitoring.
This can be explained only if spots are coming back to similar longitudes and latitudes after one stellar rotation period.
Interestingly, occulted stellar spots can place constraints on the spin-orbit alignment  (e.g \citealt{deming11,nutzman11,sanchis11a,sanchis11b}).

We identified five spots (A, B, C, D and E) by their anomaly in the residuals and we marked their initial orbital phase positions as seen in Figure~\ref{fig:spotseq}.
We then computed the expected phase position of each of these 5 anomalies on the 8 following lightcurve residuals assuming that occulted starspots should have moved by $360/8=45\degr$ in longitude towards the egress (for a prograde and aligned orbit) for every planetary transit. 
We find that anomalies are always present at all the expected phases.

As the star rotates, the spot surfaces projected on the stellar sphere change; therefore the shapes and amplitudes of these anomalies change between consecutive transits while the same spots are crossed over time. Since a spot surface appears larger at the center of the star compared to the limb, the anomalies due to spots should exhibit larger amplitudes when they are crossed close to the mid-transit time. 
This is consistent with what is observed in \koi. For example, anomaly `B' in Figure~\ref{fig:spotseq} is observed in four consecutive transits. The overall shape of this anolmaly is conserved transit after transit and its amplitude increases before and decreases after the mid-transit.

We conclude that the planet crosses the same spots, transit after transit, at longitudinal positions that differ by 45~\degr.

\begin{figure*}[h!]
\begin{center}
\includegraphics[height=5in]{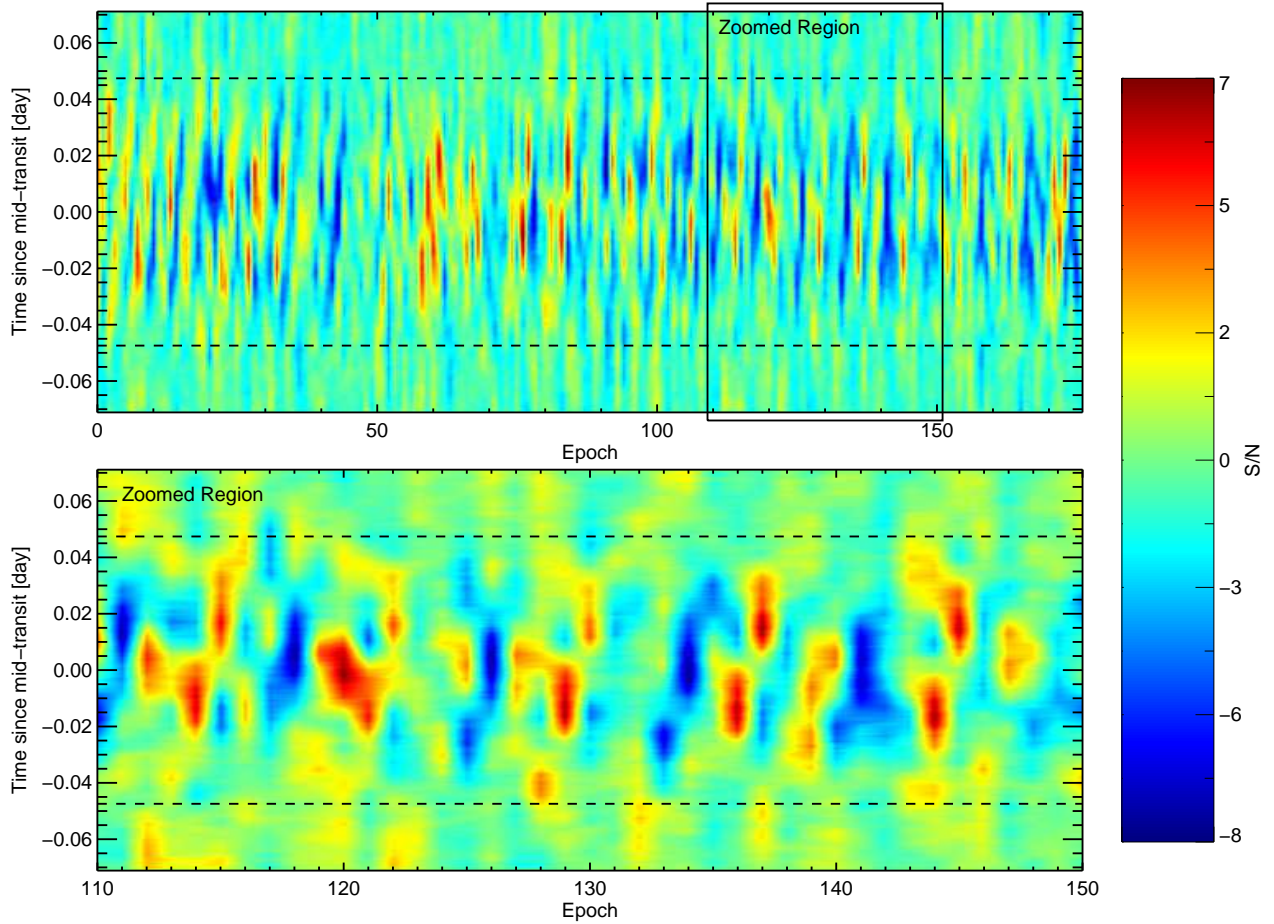} 
 \caption{Lifetime of stellar spots occulted during transits. The top panel correspond to the complete map constructed from Quarter 1 to 6 data. The bottom panel is a zoom of the top panel.
 This time-map image corresponds to the scatter (S/N) measured in a sliding box of duration twice the ingress time across the residuals in each transit epoch, relative to mid-transit (Y-axis) as a function of the epoch (X-axis). The regions in blue correspond to times when the sections of individual transit lightcurves are deeper than the averaged transit lightcurve, whereas the regions in red correspond to less deep sections attributed to cold stellar spots.
For display purposes, we used an interpolation with an output sampling of 600 samples in the epochal direction and 300 samples in the transit phase direction. }
  \label{fig:spots}
\end{center}
\end{figure*}

These findings imply that the projected spin-orbit angle, $\lambda$, is very close to 0 for this system.
Because occulted spots can be identified transit after transit (Figure~\ref{fig:spots}), with a change in phase expected from the stellar rotation period, we conclude that the stellar inclination angle is near 90 degrees. 
This is further supported by the measured $\vsini=\koicurSMEvsin$~km/s which is in good agreement with its expected value of $2\pi\rstar/P_{\rm rot} = 4.3$~km/s assuming an inclination angle of 90 degrees. 
This implies that the true obliquity, i.e. the angle between the stellar rotational axis and a line perpendicular to planet's orbital plane, $\Psi$, must be close to 0.
Finally, we conclude that the planet orbit is prograde since the occulted starspots progress from the transit ingress to the egress limbs.

Assuming that the uncertainty on latitude of a spot corresponds to the planet size, we derive an uncertainty for $\Psi$ of $\pm\arctan(\rpl/\rstar)=\pm10\degr$.  
We note that the spot size could be larger than the planet size, but this changes the results only modestly. Assuming a star spot twice the projected size of the planet on the stellar photosphere provides a more conservative uncertainty of $\pm15\degr$ for $\Psi$.

\cite{winn10} show that stars with transiting planets for which the stellar obliquity is large are preferentially hot (\teff $> 6250$~K). 
The low obliquity of \koib\ fits this pattern since the stellar temperature is lower than 6250~K.

\subsubsection{Starspots lifetime}
\label{spotlife}

To help visualize the spot crossings during transit, we subtract the best-fit model from each individual transit and investigate the residuals.  
We slide a box of duration twice the ingress time across the residuals in each transit epoch, recording the scatter for a given box mid-time (relative to mid-transit) and a given epoch.  
We then produce an image of this scatter at orbital phase as a function of the epoch number (see Figure~\ref{fig:spots}).
We used an interpolation (with IDL's TRIGRID) with an output sampling of 600 samples in the epochal direction and 300 samples in the transit phase direction.
This image reveals individual spots that we define as either ``hot" or ``cold" regions, depending on whether the individual slide box residuals are below or above the transit light curve model. 
The repeating vertical structure is interpreted as spots marching across the transit chord such as seen in the previous section.  
Each vertical profile is slanted slightly from left to right indicating that the spots progress from the ingress limb to the egress limb. 
Some spots make their way around the star and reappear again during several stellar rotation periods. 
For example, the collection of ``cold" spots in the image starting around Epoch 110 and ending around Epoch 170 seem to be related to the same spot.  
We conclude from the nearly continuous monitoring of \koi\ that the occulted starspots are present on the same stellar chord for at least 100 days, somewhat comparable to the lifetime of sunspots.

As \kepler\ continues to monitor transits of hot-Jupiters in front of active stars, it will help to better understand the stellar cycles. If the \kepler\ mission is extended, the long term photometry will enable it to produce star-spot maps and learn more about spot mean lifetimes and photospheric differential rotations. 
In the case of \koi, we may be able to measure the complete activity cycle for this star and to compare it to another well-know G-dwarf: the Sun.

\subsubsection{Impact of stellar variability of \koi\ on the system parameters}

\begin{figure*}[h!]
\begin{center}
 \includegraphics[width=5.5in]{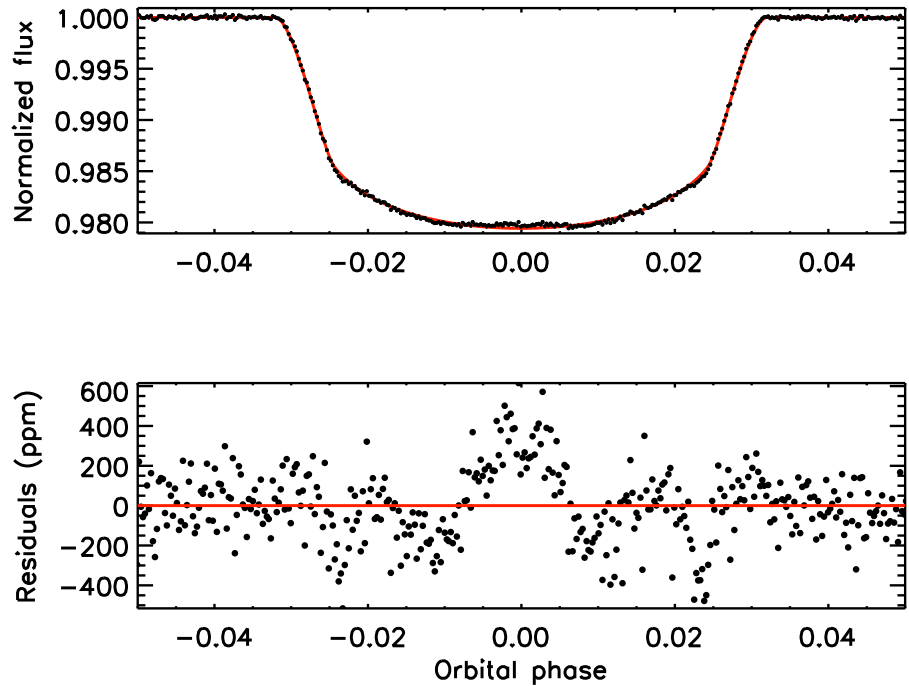}
 \caption{ Top panel: \kepler\ transit lightcurve of \koib\ phase-folded, normalized and binned by 30 seconds. The best-fit model of this lightcurve is overplotted in red. The residuals from the best-fit transit lightcurve are plotted in the bottom panel.
Occulted stellar spots produce a symmetrical pattern in phase in the residuals (see Section~\ref{starspots}).}
\label{fig:keplerlc}
\end{center}
\end{figure*}

\begin{figure*}[h!]
\begin{center}
\centering
\mbox{\subfigure{\includegraphics[width=3.5in]{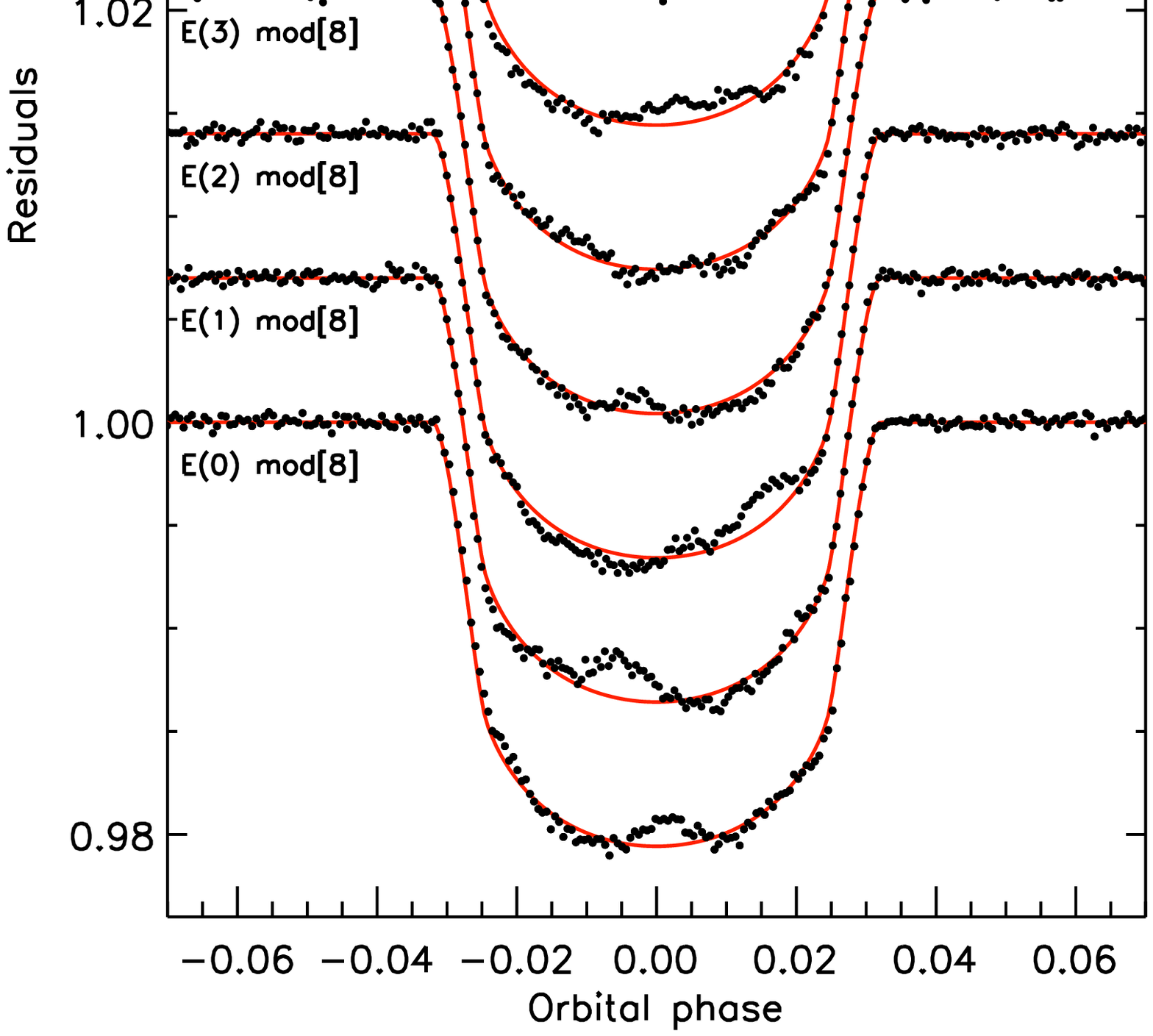}}\quad
\subfigure{\includegraphics[width=3.5in]{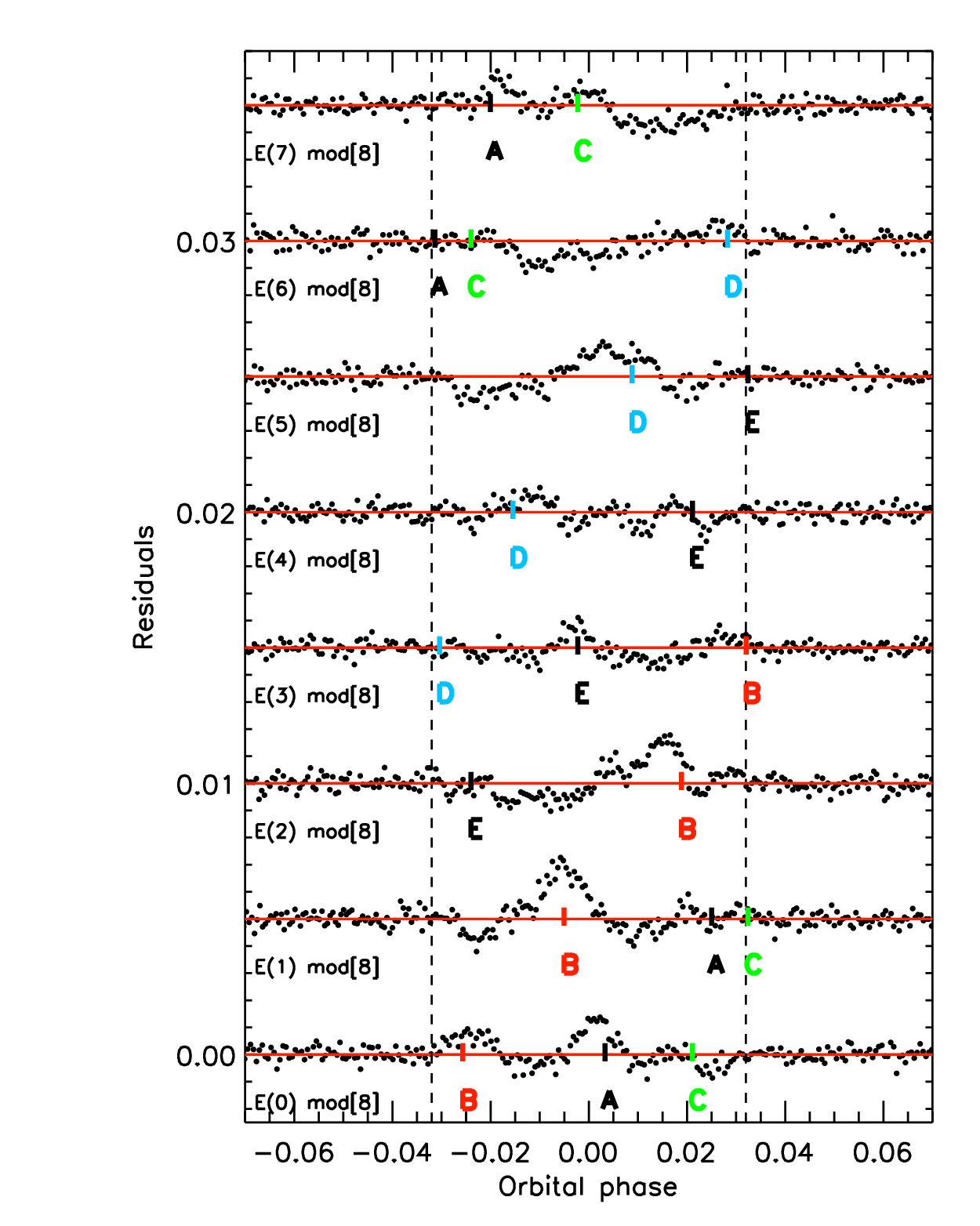}} }
\caption{{\bf Left:} Sequence of combined and binned transit lightcurves, with the best-fit model presented in Figure~\ref{fig:keplerlc} and overplotted in red. Each co-added transit corresponds to the combination of 22 individual transits that occurred at epochs modulo eight planetary orbital periods. The lightcurves are binned by 100 seconds and they are shifted vertically for display purposes. 
Each combination of individual transits allows us to increase the SNR and to demonstrate that the same spots are occulted during several consecutives transits and epochs.
The overall combination of these eight transit lightcurves gives the final curves presented in Figure~\ref{fig:keplerlc}. Occulted stellar spots are revealed in the combined curves since the stellar rotation period is eight times the planet's orbital period. The same spots are crossed every eight transits at a similar orbital phase. 
{\bf Right:} Residuals of the best-fit model subtracted from each individual combined lightcurve modulo 8. The vertical dashed lines correspond to the beginning and to the end of the transits. Five occulted stellar spots are indicated on the residuals (A, B, C, D and E) as they appear transits after transits at phase positions expected from the stellar rotation period.
This implies that the projected spin-orbit angle, $\lambda$, is very close to 0 for this system. 
The combination of the residuals of the eight transit lightcurves is similar to the total residuals plotted in Figure~\ref{fig:keplerlc} and exhibits a symmetrical structure.} 
 
\label{fig:spotseq}
\end{center}
\end{figure*}

When the planet transits in front of stellar spots, its transit shape deviates from the averaged phase-folded lightcurve. 
The effect of occulted stellar spots on the shape of the transit lightcurve is observed in the residuals from the best-fit transit model of the phase-folded light curve (see Figure~\ref{fig:keplerlc}). 
Since the stellar activity influences the transit lightcurve profiles, the planetary parameters we derive from these profiles are likely to be affected.
This is a well known problem for planets transiting in front of variable stars (e.g. \citealt{czesla09,desert11a}).
Importantly for the present study, the variability affects the stellar density that we assume a fixed value for our determination of the stellar parameters (see Section~\ref{sec:stellar}).
\cite{czesla09} propose to fit the lower envelope of the transit lightcurve to recover more realistic transit parameters.
This assumes that dark stellar structures dominate over bright faculae.
In the case of \koib, we cannot exclude the possibility that every transit is affected by dark or bright stellar regions so that a priori no individual transit light curve can be used as representative of an unaffected profile.
Furthermore, because of the stroboscopic effect described above, the phase-folded transit lightcurve possesses combined pattern distortions that prevents the use of its lower envelope to derive more accurate parameters.

We note that the combination of the residuals for the eight transit lightcurves presented in  Figure~\ref{fig:spotseq} is very similar to the total residuals plotted in Figure~\ref{fig:keplerlc} and exhibits a symmetrical structure. 
This symmetrical pattern has a shape which is consistent with what would be expected from an oblate planet but with an amplitude ten to a hundred times larger than expected for a Jupiter size planet \citep{seager02,barnes03,carter10}.
This pattern represents a remarkable coincidence, in that it seems quite symmetric but it is only obtained when all 8 time series are combined, and does not appear in the residuals for any single time series.

\subsection{Atmospheric constraints for \koib\ }\label{atmo}

The occultation depths measured in each bandpass are combined and turned into an emergent spectrum for the planet.
The observed flux of the planet in each bandpass corresponds to the sum of the reflected light and the thermally emitted light.
The combined occultation observed in the \kepler\ bandpass provides a measure of the planet's geometric albedo. 
The occultation depth of  $58 \pm 8$ ppm corresponds to $A_g=0.10 \pm 0.02$.
Assuming a Lambertian criterion $A_{\rm B} \leq 1.5 \times A_{\rm g}$, and assuming that the geometric albedo measured in the \kepler\ bandpass is the same at every wavelength, we infer an upper limit to the Bond albedo of $A_{\rm B} \leq 0.18$ at the 1-$\sigma$ level.
This result is conservative in the sense that the true albedo may be even lower, if some of the occultation signal we have measured in the \kepler\ bandpass is due to thermal emission rather than reflected light.
This is expected for the temperature regime of hot-Jupiters. 
Therefore, the Bond albedo could be well below 0.18.

We estimate the thermal component of the planet's emission in the two \spitzer\ band-passes from the occultation depths measured at 3.6 and 4.5~\micron\ (see Table~\ref{tab:spitzer}). We assume that the planetary emission is well reproduced by a black-body spectrum and translate the measured depth of the secondary eclipse into brightness temperatures. We use the PHOENIX atmospheric code \citep{hauschildt99} to produce theoretical stellar models for the star \koi. Taking the \spitzer\ spectral response function into account, the ratio of areas of the star and the planet and the stellar spectra, we derive the brightness temperatures that best-fit the observed eclipse depths measured in the two IRAC bandpasses. 
The brightness temperature calculated this way result in T$_{\rm{3.6~\micron}}$=$1880\pm110$~K and 
T$_{\rm{4.5~\micron}}$=$1770\pm150$~K.

We compare our data to the hot-Jupiter atmospheric model described in \cite{fortney08} which has been used for a variety of close-in planets \citep{fortney05,fortney06}.
We aim at broadly distinguishing between different classes of model atmospheres, given that there are no prior constraints on the basic composition and structure.
The atmospheric spectrum calculations are performed for 1D atmospheric pressure-temperature (P-T) profiles and use the equilibrium chemical abundances, at solar metallicity, described in \cite{lodders02, lodders06}. This is a self-consistent treatment of radiative transfer and chemical equilibrium of neutral species. 
The opacity database is described in \cite{freedman08}. 
The models are calculated for various values of the dayside-to-nightside energy redistribution parameter ($f$) and allow for the presence of TiO at high altitude, which may play the role of an absorber and which likely lead to an inversion of the P-T profile.

We compare the data to model predictions and select models with the best reduced \chisq. We note that this is not a fit involving adjustable parameters. 
All models shown assume redistribution of absorbed flux over the dayside only, favoring low values for $f$.
Since our observations are only weakly constraining, the 3.6 and 4.5 $\mu$m depth ratios are best fitted by a model without a temperature inversion, although the model with a temperature inversion is a reasonable fit as well. 
In the model with no inversion the optical opacity sources are neutral atomic Na and K, while in the inverted model the optical opacity is dominated by gaseous TiO and VO molecules. 
 The relatively large occultation depth in the Kepler bandpass is not matched by these simple models.  This may indicate that silicate clouds (as is seen in L-type brown dwarfs, but not modeled here) may lead to a higher amount of scattered flux from the planet.  
It may also indicate that optical opacity sources are overestimated, such that more planetary thermal flux is being seen from deeper layers.  
The growing sample size of Kepler occultation detections may show trends with planetary \teff\ that could help to clarify this issue.

\cite{hartman10} shows that there is a correlation between the surface gravity of hot Jupiters and the activity levels of the host stars, such that high surface gravity planets tend to be found around high-activity stars. 
The position of \koib\ on such a diagram is consistent with this trend.
\cite{knutson10} show that there could also be a correlation between the host star activity level and the thermal inversion of the planetary atmosphere. In this picture, the strong XEUV irradiation from the active stellar host of a hot-Jupiter depletes the atmosphere of chemical species responsible for producing inversions. The \logrhk\ of \koi\ is consistent with the case of no thermal inversion in the framework developed by \cite{knutson10}. We also evaluated the empirical index defined by \citet{knutson10}, which could be correlated with the presence of a thermal inversion. Using the same definition, we find indices of 0.026\% above and 0.034\% below the predicted black-body fluxes in the 3.6 and 4.5 micron bands. These indices suggest that the atmosphere of \koib\ is consistent with a non-inverted profile, which is in agreement with the results of our present study.

\acknowledgments

This work is based on observations made with \kepler, which was competitively selected as the tenth Discovery mission. Funding for this mission is provided by NASA's Science Mission Directorate. The authors would like to thank the many people who generously gave so much their time to make this Mission a success.

This work is also based on observations made with the \spitzer\ {\it Space Telescope},
which is operated by the Jet Propulsion Laboratory, California Institute
of Technology under a contract with NASA. Support for this work was provided by NASA through an award issued by JPL/Caltech. 

Some of the data presented herein were obtained at the W.M. Keck Observatory, which is operated as a scientific partnership among the California Institute of Technology, the University of California and the National Aeronautics and Space Administration. The Observatory was made possible by the generous financial support of the W.M. Keck Foundation.

This work is also based on observations obtained with the Hobby-Eberly Telescope (HET), which is a joint project of the University of Texas at Austin, the Pennsylvania State University, Stanford University, Ludwig-Maximilians-Universitat Munchen, and Georg-August-Universitat Gottingen. The HET is named in honor of its principal benefactors, William P. Hobby and Robert E. Eberly.

We would like to thank the Spitzer staff at IPAC and in particular Nancy Silbermann for scheduling the Spitzer observations of this program.

M. G. is a FNRS Research Associate.

\newpage


\begin{figure*}[h!]
\begin{center}
 \includegraphics[width=5.5in]{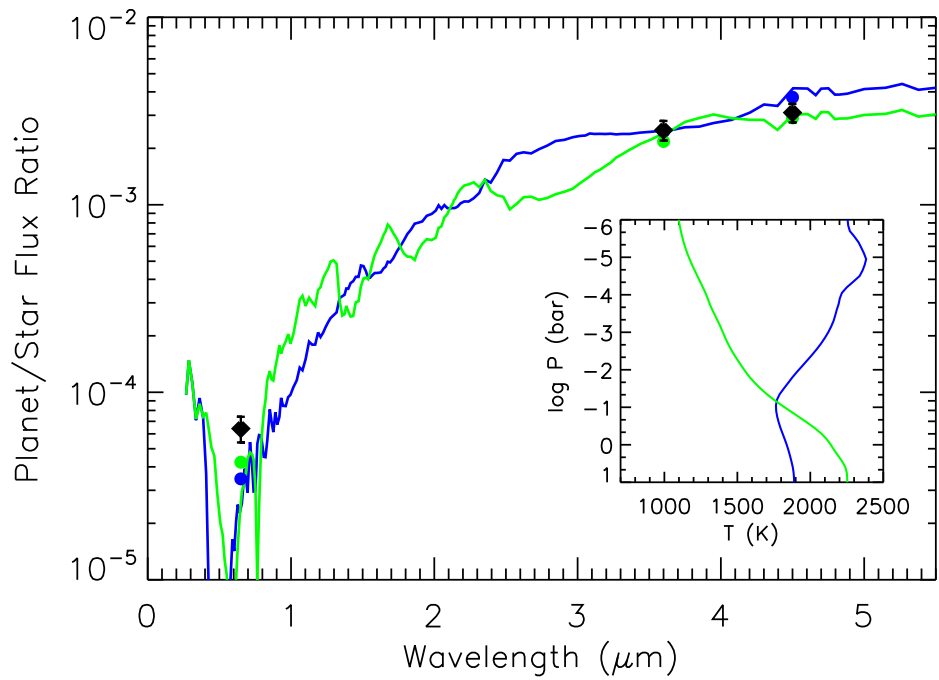}
\caption{Day side planet-to-star flux ratios as function of the wavelength for two atmospheric models \citep{fortney08}. The black filled diamonds and their error bars correspond to the \kepler\ and \spitzer\ observations. The dataset is best fitted with models assuming full redistribution of the energy to the planetary day-side. The green model represent a non-inverted atmosphere (no TiO). The blue model represents an inverted atmosphere (with TiO). The target is slightly better fitted by the non-inverted model.}
   \label{fig:jonmodels}
\end{center}
\end{figure*}



\newpage

\begin{deluxetable}{rrrrrr}
\tablewidth{0pc}
\tablecaption{Relative Radial-Velocity Measurements of \koi{}
\label{tab:RVs}}
\tablehead{
\colhead{HJD}                           &
\colhead{Phase}                         &
\colhead{RV}                            &
\colhead{\ensuremath{\pm \sigma_{\rm RV}}}  &
\colhead{BS}                            &
\colhead{\ensuremath{\pm \sigma_{\rm BS}}}  \\
\colhead{(days)}                        &
\colhead{(cycles)}                      &
\colhead{(\ms)}                         &
\colhead{(\ms)}                         &
\colhead{(\ms)}                         &
\colhead{(\ms)}
}
\startdata

   2455430.839929  & 313.013 &$    -142.61 $&$   67.08 $&$ -77.7    $&$ 114.6   $\\
   2455431.812259  & 313.667 &$     325.71  $&$ 49.17 $&$  -23.8    $&$ 72.4 $\\
   2455432.819992  & 314.345 &$    -300.32 $&$   90.74 $&$  42.2     $&$ 68.9 $\\
   2455442.794995  & 321.060 &$    -119.08 $&$   43.22 $&$  45.6     $&$ 68.2$\\
   2455443.765027  & 325.080 &$     528.51  $&$ 198.75 $&$  483.3   $&$ 276.7$\\
   2455448.768015  & 327.760 &$    -158.75 $&$   51.43 $&$  -110.9 $&$  218.2$\\
   2455452.748541  & 341.195 &$     358.36  $&$  61.44 $&$  -50.3    $&$  67.7$\\
   2455472.707059  & 346.565 &$    -413.96 $&$   25.96 $&$  -113.8 $&$ 80.9$\\
   2455480.686032  & 355.957 &$     138.23  $&$  56.24 $&$  -116.1  $&$ 70.4$\\
   2455494.639064  & 357.302 &$      83.68   $&$ 36.83 $&$  -47.9     $&$ 75.7$\\
   2455496.636013  & 372.077 &$    -511.48 $&$   60.28 $&$  -36.6   $&$  75.3$\\
   2455518.588037  & 372.747 &$    -205.45 $&$   56.70 $&$  1.2       $&$ 67.6$\\
   2455522.579531  & 374.764 &$     417.16  $&$  28.10 $&$ -121.9   $&$ 34.2 $\\

\enddata
\end{deluxetable}

\begin{center}
\begin{deluxetable*}{lccccccc}
\tabletypesize{\scriptsize}
\tablecaption{{\it Warm-Spitzer} observations. }
\tablewidth{0pt}
\tablehead{\colhead{Visit} & \colhead{AOR} & \colhead{Wavelength (\micron)} & \colhead{Obs. Date (UT)} & \colhead{Select. points} & \colhead{Depth (\%)} & \colhead{Weighted. Avg. depth} & \colhead{Bright. T (K)}}
\startdata
1 & 40252416 & 3.6  &  2010-08-29   & 2266 & $0.311^{+0.045}_{-0.060}$ &              -                  &  \\
3 & 40251904 & 3.6  &  2010-09-16   & 2301 & $0.235^{+0.030}_{-0.030}$ & $0.250\pm0.030$\% & $1880\pm110$ \\
2 & 40252160 & 4.5  &  2010-08-30   & 2315 & $0.270^{+0.040}_{-0.060}$ &              -                   & \\
4 & 40251648 & 4.5  &  2010-09-18   & 2313 & $0.346^{+0.060}_{-0.050}$ & $0.310\pm0.035$\% &   $1770\pm150$\\
\enddata
\label{tab:spitzer}
\end{deluxetable*}
\end{center}

\begin{center}
\begin{deluxetable*}{lcc}
\tabletypesize{\scriptsize}
\tablewidth{0pc}
\tablecaption{System Parameters for \koi \label{tab:param}}
\tablehead{\colhead{Parameter}	& 
\colhead{Value} 		& 
\colhead{Notes}}
\startdata
\sidehead{\em Object of interest}
Kepler Input Catalog number (KIC) 				& $10619192$		&\\
Kepler Object of Interest number (KOI)				& $203$		&	\\
RA (J2000)  				&  $19:53:34.86$ 		& 	\\
Dec (J2000) 				& $+47:48:54.0$		&	\\
Kepler magnitude    & $14.14$		&	\\
J magnitude             & $12.99$		&	\\
r magnitude             & $14.08$		&	\\

\sidehead{\em Transit and orbital parameters}
Orbital period $P$ (d)				& \koicurLCP		& A	\\
Transit Ephemeris (BJD$_{utc}$)			& \koicurLCT		& A	\\
Scaled semimajor axis $a/\rstar$		& \koicurLCar		& A	\\
Scaled planet radius \rpl/\rstar		& \koicurLCrprstar	& A	\\
Impact parameter $b \equiv a \cos{i}/\rstar$	& \koicurLCimp		& A	\\
Orbital inclination $i$ (deg)			& \koicurLCi 		& A	\\
Orbital semi-amplitude $K$ (\ms)		& \koicurRVK		& A,B	\\
$\sqrt{e}cos(w)$		&	  ${\ensuremath{0.008^{+0.015}_{-0.013}}}$ & A,B	\\
$\sqrt{e}sin(w)$	    &	  ${\ensuremath{-0.084^{+0.027}_{-0.033}}}$ & A,B	\\
Orbital eccentricity $e$			&  $<0.011$		& A,B	\\
Transit duration (d) & $0.09485^{+0.00007}_{-0.00007}$ \\
\sidehead{\em Observed stellar parameters (fixing the surface gravity)}
Effective temperature \teff\ (K)		& \koicurSMEteff	& C 	\\
Metallicity \mh				& \koicurSMEfeh		& C	\\
Projected rotation \vsini\ (\kms)		& \koicurSMEvsin	& C	\\
Absolute Systemic Radial Velocity $\gamma$  (\kms)		& $-23.82\pm0.10$ & C	\\
Mean radial velocity (\kms)			& \koicurRVmean		& B	\\
Quadratic limb darkening coefficient (u$_{1}$)				& \koicurLCua		& A	\\
Quadratic limb darkening coefficient (u$_{2}$)				& \koicurLCub		& A	\\
\sidehead{\em Derived stellar parameters}
Mass \mstar (\msun)				& \koicurYYmlong	& C,D	\\
Radius \rstar (\rsun)  				& \koicurYYrlong	& C,D	\\
Surface gravity \loggstar\ (cgs)		& \koicurYYlogg		& A	\\
Age (Gyr)					& \koicurYYage		& C,D	\\
\sidehead{\em Planetary parameters}
Mass \mpl\ (\mjup)				& \koicurPPm		& A,B,C,D	\\
Radius \rpl\ (\rjup, equatorial)		& \koicurPPr		& A,B,C,D	\\
Density \rhopl\ (\gcmc)				& \koicurPPrho		& A,B,C,D	\\
Surface gravity \loggpl\ (cgs)			& \koicurPPlogg		& A,B	\\
Orbital semimajor axis $a$ (AU)			& \koicurPParel		& E	\\
Equilibrium temperature \teq\ (K)		& \koicurPPteq		& F	\\
Geometric albedo A$_{g}$			&  $0.10\pm0.02$		& A	\\
\enddata
\tablecomments{\\
A: Based on the photometry.\\
B: Based on the radial velocities.\\
C: Based on an analysis of the TRES spectrum.\\
D: Based on the Yale-Yonsei stellar evolution tracks.\\
E: Based on Newton's version of Kepler's Third Law and total mass.\\
F: Assumes Bond albedo = 0. and complete redistribution.
}
\end{deluxetable*}
\end{center}



\clearpage

\newpage

\begin{thebibliography}{}

\bibitem[Agol et al.(2005)]{agol05} Agol, E., Steffen, J., Sari, R., \& Clarkson, W.\ 2005, \mnras, 359, 567 
\bibitem[Alonso et al.(2010)]{alonso10} Alonso, R., Deeg, H.~J., Kabath, P., \& Rabus, M.\ 2010, \aj, 139, 1481 
\bibitem[Alonso et al.(2009a)]{alonso09a} Alonso, R., Guillot, T., Mazeh, T., Aigrain, S., Alapini, A., Barge, P., Hatzes, A., \& Pont, F.\ 2009, \aap, 501, L23 
\bibitem[Anderson et al.(2011)]{Anderson2011} Anderson, D.~R., et al.\ 2011, \apjl, 726, L19 
\bibitem[Barnes \& Fortney(2003)]{barnes03} Barnes, J.~W., \& Fortney, J.~J.\ 2003, \apj, 588, 545 
\bibitem[Barman et al.(2005)]{barman05} Barman, T.~S., Hauschildt, P.~H., \& Allard, F.\ 2005, \apj, 632, 1132 
\bibitem[Batalha et al.(2010)]{batalha10} Batalha, N.~M., et al.\ 2010, \apjl, 713, L109 
\bibitem[Borucki et al.(2009)]{borucki09} Borucki, W. J. et al. 2009, Science, 325, 709
\bibitem[Borucki et al.(2010)]{borucki10} Borucki, W.~J., et al.\ 2010, Science, 327, 977
\bibitem[Borucki et al.(2011)]{borucki11} Borucki, W.~J., et al.\ 2011, \apj, 736, 19
\bibitem[Burrows et al.(2005)]{burrows05} Burrows, A., Hubeny, I., \& Sudarsky, D.\ 2005, \apjl, 625, L135 
2007, \apj, 668, L171
\bibitem[Burrows et al.(2008)]{burrows08} Burrows, A., Ibgui, L., \& Hubeny, I.\ 2008, \apj, 682, 1277 
\bibitem[Brown et al.(2011)]{brown11} Brown, T.~M., Latham, D.~W., Everett, M.~E., \& Esquerdo, G.~A.\ 2011, arXiv:1102.0342 
\bibitem[Bryson et al.(2010)]{bryson10} Bryson, S.~T., et al. 2010, \apjl, 713, L97 
\bibitem[Buchhave et al.(2010)]{buchhave10} Buchhave, L.~A., et al.~2010, ApJ, 720, 1118
\bibitem[Caldwell et al.(2010)]{caldwell10} Caldwell, D.~A., et al.\ 2010, \apjl, 713, L92
\bibitem[Carter et al.(2011)]{carter11} Carter, J.~A., Winn,  J.~N., Holman, M.~J., Fabrycky, D., Berta, Z.~K., Burke, C.~J., 
\& Nutzman, P.\ 2011, \apj, 730, 82 

\bibitem[Carter \& Winn(2010)]{carter10} Carter, J.~A., \& Winn, J.~N.\ 2010, \apj, 709, 1219 

\bibitem[Charbonneau et al.(2005)]{charbonneau05} Charbonneau, D., et al. 2005, \apj, 626, 523
\bibitem[Czesla et al.(2009)]{czesla09} Czesla, S., Huber, K.~F., Wolter, U., Schr{\"o}ter, S., \& Schmitt, J.~H.~M.~M.\ 2009, \aap, 505, 1277 

\bibitem[Deming et al.(2011)]{deming11} Deming, D., et al.\ 2011, arXiv:1107.2977, \apj, submitted 
\bibitem[Deming et al.(2005)]{deming05} Deming, D., Seager, S., \& Richardson, L. J. 2005, \nat, 434, 740
\bibitem[Demarque et al.(2004)]{demarque04} Demarque, P., Woo, J.-H., Kim, Y.-C., \& Yi, S.~K.\ 2004, \apjs, 155, 667 

\bibitem[D{\'e}sert et al.(2011b)]{desert11b} D{\'e}sert, J.-M., et al.\ 2011, arXiv:1102.0555, \apj, submitted
\bibitem[D{\'e}sert et al.(2011a)]{desert11a} D{\'e}sert, J.-M., et al.\ 2011, \aap, 526, A12 


\bibitem[D{\'e}sert et al.(2009)]{desert09} D{\'e}sert, J.-M., Lecavelier des Etangs, A., H{\'e}brard, G., Sing, D.~K., Ehrenreich, D., Ferlet, R., \& Vidal-Madjar, A.\ 2009, \apj, 699, 478 


\bibitem[Endl et al.(2011)]{endl11}Endl, M., et al.\ 2011, arXiv:1107.2596, \apj, submitted 

\bibitem[Fortney et al.(2005)]{fortney05} Fortney, J.~J., Marley, M.~S., Lodders, K., Saumon, D., \& Freedman, R.\ 2005, \apjl, 627, L69 
\bibitem[Fortney et al.(2006)]{fortney06} Fortney, J. J. et al. 2006, \apj, 652, 746
\bibitem[Fortney et al.(2008)]{fortney08} Fortney, J. J., Lodders, K., Marley, M. S., \& Freedman, R. S. 2008, \apj, 678, 1419
\bibitem[Freedman et al.(2008)]{freedman08} Freedman, R.~S., Marley, M.~S., \& Lodders, K.\ 2008, \apjs, 174, 504 
\bibitem[Fazio et al.(2004)]{fazio04} Fazio, G.~G., et al. 2004, \apjs, 154, 10
\bibitem[F\H{u}r\'esz(2008)]{furesz08} F\H{u}r\'esz, G.~2008, Ph.D.~thesis, University of Szeged, Hungary
\bibitem[Gautier et al.(2010)]{gautier10} Gautier, T.~N., III, et al.\ 2010, arXiv:1001.0352 
\bibitem[Gilliland et al.(2010)]{gilliland10} Gilliland, R.~L., et al.\ 2010, \apjl, 713, L160 
\bibitem[Gillon et al.(2010)]{gillon10} Gillon, M., et al. 2010, \aap, 511, A3
\bibitem[Gillon et al.(2009)]{gillon09} Gillon, M., et al.\ 2009, \aap, 506, 359 
249
\bibitem[Hartman(2010)]{hartman10} Hartman, J.~D.\ 2010, \apjl, 717, L138 
\bibitem[Hauschildt et al.(1999)]{hauschildt99} Hauschildt, P. H., Allard, F., \& Baron, E. 1999, \apj, 512, 377
\bibitem[Holman et al.(2006)]{holman06} Holman, M.~J., et al.\ 2006, \apj, 652, 1715 
\bibitem[Holman \& Murray(2005)]{holman05} Holman, M.~J., \& Murray, N.~W.\ 2005, Science, 307, 1288 

\bibitem[Isaacson \& Fischer(2010)]{isaacson10} Isaacson, H., \& Fischer, D.\ 2010, \apj, 725, 875 
\bibitem[Jenkins et al.(2010)]{jenkins10b} Jenkins, J.~M., et al. 2010, \apjl, 713, L120 
\bibitem[Jenkins et al.(2010)]{jenkins10a} Jenkins, J.~M., et al. 2010, \apjl, 713, L87 
\bibitem[Jefferys et al.(1988)]{jefferys88} Jefferys, W.~H., Fitzpatrick, M.~J., \& McArthur, B.~E.\ 1988, Celestial Mechanics, 41, 39 
submitted to ApJ 
\bibitem[Knutson et al.(2008)]{knutson08} Knutson, H. A. et al. 2008, \apj, 673, 526
\bibitem[Knutson et al.(2010)]{knutson10} Knutson, H.~A., Howard, A.~W., \& Isaacson, H.\ 2010, \apj, 720, 1569 
\bibitem[Koch et al.(2010)]{koch10b} Koch, D.~G., et al.\ 2010, \apjl, 713, L131 
\bibitem[Koch et al.(2010)]{koch10a} Koch, D.~G., et al.\ 2010, \apjl, 713, L79 
\bibitem[Kurucz(1979)]{kurucz79} Kurucz, R.~L.\ 1979, \apjs, 40, 1 
\bibitem[Latham et al.(2010)]{latham10} Latham, D.~W., et al. 2010, \apjl, 713, L140 
\bibitem[Lodders \& Fegley(2006)]{lodders06} Lodders, K., \& Fegley, B., Jr.\ 2006, Astrophysics Update 2,1 
\bibitem[Lodders(2002)]{lodders02} Lodders, K.\ 2002, \apj, 577, 974 
\bibitem[Nidever et al.(2002)]{nidever02} Nidever, D.~L., et al.~2002, ApJ, 141, 503

\bibitem[Nutzman et al.(2011)]{nutzman11} Nutzman, P.~A., Fabrycky, D.~C., \& Fortney, J.~J.\ 2011, arXiv:1107.2106, \apj, submitted
\bibitem[Press et al.(1992)]{press92} Press, W.~H., Teukolsky, S.~A., Vetterling, W.~T., \& Flannery, B.~P.\ 1992, Cambridge: University Press, c1992, 2nd ed.,  
\bibitem[Queloz et al.(2001)]{queloz01} Queloz, D., et al.\ 2001, \aap, 379, 279 
\bibitem[Reach et al.(2006)]{reach06} Reach, W.~T., et al. 2006, IRAC Data Handbook v3.0   
\bibitem[Rowe et al.(2008)]{rowe08} Rowe, J.~F., et al.\ 2008, \apj, 689, 1345
\bibitem[Rowe et al.(2006)]{rowe06} Rowe, J.~F., et al.\ 2006, \apj, 646, 1241 
\bibitem[Sanchis-Ojeda \& Winn(2011b)]{sanchis11b} Sanchis-Ojeda, R., \& Winn, J.~N.\ 2011, arXiv:1107.2920, \apj, submitted
\bibitem[Sanchis-Ojeda et al.(2011a)]{sanchis11a} Sanchis-Ojeda,  R., Winn, J.~N., Holman, M.~J., Carter, J.~A., Osip, D.~J.,  \& Fuentes, C.~I.\ 2011, \apj, 733, 127
\bibitem[Seager et al.(2000)]{seager00} Seager, S., Whitney, B.~A., \& Sasselov, D.~D.\ 2000, \apj, 540, 504 
\bibitem[Seager \& Mall{\'e}n-Ornelas(2003)]{seager03} Seager, S., \& Mall{\'e}n-Ornelas, G.\ 2003, \apj, 585, 1038 
\bibitem[Seager \& Hui(2002)]{seager02} Seager, S., \& Hui, L.\ 2002, \apj, 574, 1004 
\bibitem[Snellen et al.(2009)]{snellen09} Snellen, I. A. G., de Mooij, E. J. W., \& Albrecht, S. 2009, \nat, 459, 543
\bibitem[Southworth(2011)]{southworth11} Southworth, J.\ 2011, arXiv:1107.1235, \aa, submitted
\bibitem[Southworth et al.(2007)]{southworth07} Southworth, J., Wheatley, P.~J., \& Sams, G.\ 2007, \mnras, 379, L11
\bibitem[Sozzetti et al.(2007)]{sozzetti07} Sozzetti, A., Torres, 
G., Charbonneau, D., Latham, D.~W., Holman, M.~J., Winn, J.~N., Laird, 
J.~B., \& O'Donovan, F.~T.\ 2007, \apj, 664, 1190 
\bibitem[Spiegel \& Burrows(2010)]{spiegel10} Spiegel, D.~S., \& Burrows, A.\ 2010, \apj, 722, 871 
\bibitem[Sudarsky et al.(2000)]{sudarsky00} Sudarsky, D., Burrows, A., \& Pinto, P.\ 2000, \apj, 538, 885
\bibitem[Sudarsky et al.(2003)]{sudarsky03} Sudarsky, D., Burrows, A., \& Hubeny, I.\ 2003, \apj, 588, 1121 
\bibitem[Torres et al.(2008)]{torres08} Torres, G., Winn, J.~N., \& Holman, M.~J.\ 2008, \apj, 677, 1324
\bibitem[Tull(1998)]{tull98} Tull, R.~G.\ 1998, \procspie,  3355, 387
\bibitem[Welsh et al.(2010)]{welsh10} Welsh, W.~F., Orosz, J.~A., Seager, S., Fortney, J.~J., Jenkins, J., Rowe, J.~F., Koch, D., \& Borucki, W.~J.\ 2010, \apjl, 713, L145 
\bibitem[Werner et al.(2004)]{werner04} Werner, M.~W. et al., 2004, \apjs, 154, 1
\bibitem[Winn et al.(2010)]{winn10} Winn, J.~N., Fabrycky, D.,  Albrecht, S., \& Johnson, J.~A.\ 2010, \apjl, 718, L145 
\bibitem[Winn et al.(2008)]{winn08} Winn, J.~N., et al.\ 2008, \apj, 683, 1076 
\bibitem[Yi et al.(2001)]{yi01} Yi, S., Demarque, P., Kim,  Y.-C., Lee, Y.-W., Ree, C.~H., Lejeune, T.,  \& Barnes, S.\ 2001, \apjs, 136, 417 









\end{thebibliography}
\end{document}